\documentclass[lettersize,journal]{IEEEtran}
\usepackage{amsmath,amsfonts,amssymb,mathtools}
\usepackage{algorithmic}
\usepackage{algorithm}
\usepackage{array}
\usepackage[caption=false,font=normalsize,labelfont=sf,textfont=sf]{subfig}
\usepackage{textcomp}
\usepackage{stfloats}
\usepackage{url}
\usepackage{verbatim}
\usepackage{graphicx}
\usepackage{cite}
\usepackage[hidelinks]{hyperref}
\usepackage{tikz}
\usepackage{pgfplots}
\usepackage{amsthm}
\usepackage{booktabs}
\usepackage{etoolbox}
\usepackage{orcidlink}

\pgfplotsset{compat=1.18}

\newtheorem{theorem}{Theorem}
\newtheorem{lemma}{Lemma}

\newcommand\copyrighttext{%
  \footnotesize
  \textcopyright 2025 IEEE. Personal use of this material is permitted.  Permission from IEEE must be obtained for all other uses, in any current or future media, including reprinting/republishing this material for advertising or promotional purposes, creating new collective works, for resale or redistribution to servers or lists, or reuse of any copyrighted component of this work in other works. DOI: \href{https://doi.org/10.1109/TSP.2025.3544170}{10.1109/TSP.2025.3544170}}
\newcommand\copyrightnotice{%
\begin{tikzpicture}[remember picture,overlay]
\node[anchor=south,yshift=5pt] at (current page.south) {\fbox{\parbox{\dimexpr\textwidth-\fboxsep-\fboxrule\relax}{\copyrighttext}}};
\end{tikzpicture}%
}

\begin{document}

\title{Subspace Representation Learning for Sparse Linear Arrays to Localize More Sources than Sensors:\\A Deep Learning Methodology}

\author{\orcidlinki{Kuan-Lin Chen}{0009-0005-4067-0927},~\IEEEmembership{Member,~IEEE,} and \orcidlinki{Bhaskar D. Rao}{0000-0001-6357-689X},~\IEEEmembership{Life Fellow,~IEEE}
\thanks{The authors are with the Department of Electrical and Computer Engineering, University of California, San Diego, La Jolla, CA 92093, USA
(e-mail: \texttt{kuc029@ucsd.edu}; \texttt{brao@ucsd.edu}).}}



\maketitle
\copyrightnotice
\begin{abstract}
  Localizing more sources than sensors with a sparse linear array (SLA) has long relied on minimizing a distance between two covariance matrices and recent algorithms often utilize semidefinite programming (SDP). Although deep neural network (DNN)-based methods offer new alternatives, they still depend on covariance matrix fitting. In this paper, we develop a novel methodology that estimates the co-array subspaces from a sample covariance for SLAs. Our methodology trains a DNN to learn signal and noise subspace representations that are invariant to the selection of bases. To learn such representations, we propose loss functions that gauge the separation between the desired and the estimated subspace. In particular, we propose losses that measure the length of the shortest path between subspaces viewed on a union of Grassmannians, and prove that it is possible for a DNN to approximate signal subspaces. The computation of learning subspaces of different dimensions is accelerated by a new batch sampling strategy called consistent rank sampling. The methodology is robust to array imperfections due to its geometry-agnostic and data-driven nature. In addition, we propose a fully end-to-end gridless approach that directly learns angles to study the possibility of bypassing subspace methods. Numerical results show that learning such subspace representations is more beneficial than learning covariances or angles. It outperforms conventional SDP-based methods such as the sparse and parametric approach (SPA) and existing DNN-based covariance reconstruction methods for a wide range of signal-to-noise ratios (SNRs), snapshots, and source numbers for both perfect and imperfect arrays.
\end{abstract}


\section{Introduction}
\IEEEPARstart{D}{irection}-of-arrival (DoA) estimation is one of the fundamental problems in array processing, providing the direction information of sources to many applications such as hearing aids \cite{pisha2019wearable}, wireless communications \cite{sant2020doa}, and sonar systems \cite{liu2021doa}. When a sufficiently large number of array measurements or \textit{snapshots} are available, most approaches estimate a spatial covariance matrix (SCM) and apply subspace methods like MUtiple SIgnal Classification (MUSIC) \cite{schmidt1986multiple} to find the DoAs. Because the noise subspace is required to be nontrivial, an $M$-element uniform linear array (ULA) can only resolve up to $M-1$ sources. To remove such a limit and reduce the cost of sensors, one can choose an $N$-element SLA with the same aperture but no ``holes'' in its co-array \cite{van2002optimum}. In this case, the $M$-by-$M$ SCM of the original ULA can be reconstructed from the $N$-by-$N$ SCM of the SLA. Taking a $5$-element minimum redundancy array (MRA) for example, it can recover the SCM of a $10$-element ULA and thus resolve up to $9$ sources with only $5$ sensors. Although such an exploitation on the co-array structure can deliver more degrees of freedom, an extra step of covariance matrix estimation is required \cite{sarangi2023super}.

The earliest approach to this problem dates back to the work by Pillai \textit{et al.} in 1985, which completes a Toeplitz matrix via redundancy averaging and direct augmentation \cite{pillai1985new}. Since the SCM of a ULA is positive semidefinite and possibly Toeplitz, the matrix estimation problem can be formulated as constrained optimization problems under the well-known maximum likelihood (ML) principle. However, these problems are nontrivial due to being highly nonconvex, and one often needs to relax them into convex optimization problems. For example, the problem of the coarray ML-MUSIC (Co-MLM) \cite{qiao2017maximum} is usually relaxed into the SDP problem of SPA \cite{yang2014discretization} according to the extended invariance principle \cite{stoica1989reparametrization,ottersten1998covariance}, with its global minimizer approximating the ML estimator as the number of snapshots approaches infinity. Besides convex relaxation, another strategy to tackle nonconvex optimization is majorization-minimization. For instance, the recently proposed StructCovMLE approach by Pote and Rao \cite{pote2023maximum} majorizes the concave component by a supporting hyperplane and then solves a sequence of SDP problems to arrive at a solution. There are also many other approaches such as regularized algorithms based on nuclear norm or atomic norm minimization \cite{tang2014near,li2015off}, Wasserstein distance minimization \cite{wang2019grid}, and proxy covariance estimation \cite{sarangi2021beyond}. Literature on DoA estimation that primarily relies on optimization techniques is vast \cite{wu2017toeplitz,zhou2018direction}, so we focus on gridless and regularizer-free approaches in this paper. For grid-based DoA estimation, we refer the reader to other references such as \cite{stoica2010spice} and \cite{stoica2012spice}.

In the past decade, the advent of deep learning has opened up a new paradigm for DoA estimation \cite{liu2018direction,papageorgiou2021deep,barthelme2021machine,chen2023dnn}. As the most intuitive and earliest learning-based approach, one can discretize the angle domain into a grid and then learn a classifier \cite{papageorgiou2021deep}. However, the performance of this approach is limited by the grid size and often the performance quickly saturates as the SNR increases. On the gridless side, it was not until a recent work by Wu \textit{et al.} \cite{wu2022gridless} that the potential of deep learning for the matrix estimation problem was shown. Based on enforcing the Toeplitz structure of the matrix, they showed that DNNs can be trained to retrieve the noiseless SCM of a ULA from the sample SCM of an MRA, and numerical results show that such an approach outperforms the SPA in most cases. However, it was reported that its performance is worse than MUSIC when the source number is small at high SNRs. Another feature that makes the approach slightly less appealing is that a separate DNN is required for each individual source number. It is unknown whether training one DNN for all source numbers can still provide good performance. In contrast to using the Toeplitz structure, the framework proposed by Barthelme and Utschick \cite{barthelme2021doa} enforces the structure of positive definiteness of the matrix. Although in \cite{barthelme2021doa} the task of interest is subarray sampling, which is different from the present paper, the method can be applied seamlessly to the matrix estimation problem here. These two approaches are probably the most relevant related work to this paper.

In this paper, we propose a new methodology that exploits the fundamental property that a subspace is invariant of the choice of the spanning basis, and answer the following question: \textit{Is it possible for a neural network to learn the signal or noise subspace?} In particular, we formulate the DoA estimation problem as a subspace representation learning problem, and propose new empirical risk minimization problems and loss functions to train a DNN to learn subspace representations. Our approach first constructs a DNN to output a square matrix and performs eigenvalue decomposition on the Gram matrix of the square matrix to obtain unitary bases for the signal and noise subspaces, which we refer to as subspace representations. The DNN is then trained by minimizing loss functions of different dimensions based on principal angles that calculate the average degree of separation between the desired subspace and the subspace representation. In fact, with this new methodology, one can argue that learning subspaces is simpler than learning covariance matrices. Because our loss functions are invariant to the selection of bases, they create a larger solution space and thus make it easier for a DNN to learn subspace structures. Furthermore, we prove that it is possible for a neural network to approximate signal subspaces. To parallelize the computation of learning subspaces of different dimensions, we propose a new batch sampling strategy called consistent rank sampling, which greatly accelerates the training process. In addition, we propose a new gridless end-to-end approach learning DoAs directly to study the benefit of bypassing the root-MUSIC algorithm. Our methodology does not require knowledge of the sensor array positions, making it geometry-agnostic and robust to array imperfections. Under the standard assumptions of DoA estimation, numerical results show that our approach outperforms existing SDP-based and DNN-based methods across a wide range of SNRs, snapshots, and numbers of sources.
\section{Preliminaries} \label{sec:prelim}
\noindent Notations, assumptions, definitions, and the problem of interest are set up in this section. The set $\{1,2,\cdots,n\}$ is denoted by $[n]$. The zero-mean circularly symmetric complex Gaussian distribution with covariance $\boldsymbol{\Sigma}$ is denoted by $\mathcal{CN}(\mathbf{0},\boldsymbol{\Sigma})$. The Frobenius norm of a matrix $\mathbf{A}$ is denoted by $\lVert\mathbf{A}\rVert_F$. The trace of a matrix $\mathbf{A}$ is denoted by $\text{tr}(\mathbf{A})$. The set of $n$-by-$n$ Hermitian matrices is denoted by $\mathbb{H}^n$. Given $\mathbf{A}\in\mathbb{H}^n$, $\mathbf{A}\succeq 0$ (resp., $\mathbf{A}\succ 0$) means that $\mathbf{A}$ is positive semidefinite (resp., positive definite). The set of $n$-by-$n$ Toeplitz matrices is denoted by $\mathbb{T}^n$. For every $\mathbf{A}\in\mathbb{H}^n\cap\mathbb{T}^n$ whose first row is represented by a vector $\mathbf{u}$, $\mathbf{A}$ is denoted as $\text{Toep}(\mathbf{u})$. The minimum eigenvalue of a matrix $\mathbf{A}\succeq 0$ is denoted by $\lambda_{\text{min}}(\mathbf{A})$. The matrix logarithm of $\mathbf{A}$ is denoted by $\log\left(\mathbf{A}\right)$ \cite{higham2008functions}. The set of all $k$-by-$k$ permutation matrices is denoted by $\mathcal{P}_k$. The orthogonal projector onto a subspace $\mathcal{U}$ and the range of a matrix $\mathbf{A}$ are denoted by $P_{\mathcal{U}}$ and $P_{\mathbf{A}}$, respectively.

\subsection{Assumptions} \label{assumptions}
\noindent Let us consider an $M$-element ULA with spacing $d=\frac{\lambda}{2}$ centered at the origin. Assume that there are $k$ narrowband and far-field source signals $\{s_i\}_{i=1}^k$ with a carrier wavelength $\lambda$ impinging on the array from DoAs
$\boldsymbol{\theta}=\{\theta_1,\theta_2,\cdots,\theta_k\}\subset[0,\pi]$. Under the plane wave assumption \cite{van2002optimum}, the received array measurement vector or snapshot $\mathbf{y}(t)\in\mathbb{C}^M$ at time $t\in[T]$ can be modeled as
\begin{equation}
    \mathbf{y}(t)=\sum_{i=1}^{k}s_i(t)\mathbf{a}(\theta_i)+\mathbf{n}(t)=\mathbf{A}(\boldsymbol{\theta})\mathbf{s}(t)+\mathbf{n}(t)
\end{equation}
where $\mathbf{a}(\theta):[0,\pi]\to\mathbb{C}^M$ is the array manifold of the $M$-element ULA whose $i$-th element is given by
\begin{equation}
    [\mathbf{a}(\theta)]_i=e^{j2\pi \left(i-1-\frac{(M-1)}{2}\right)\frac{d}{\lambda}\cos\theta}, i\in[M]
\end{equation}
and $\mathbf{A}(\boldsymbol{\theta})=\begin{bmatrix}\mathbf{a}(\theta_1)&\mathbf{a}(\theta_2)&\cdots&\mathbf{a}(\theta_k)\end{bmatrix}$. The source signal vectors are given by $\mathbf{s}(t)=\begin{bmatrix}s_1(t)&s_2(t)&\cdots&s_k(t)\end{bmatrix}^{\mathsf{T}}$ for all $t\in[T]$ and are independent and identically distributed (i.i.d.) with $\mathbf{s}(t)\sim\mathcal{CN}\left(\mathbf{0},\mathbf{P}\right)$ where $\mathbf{P}=\text{diag}(p_{1},p_{2},\cdots,p_{k})$ and $p_i>0$ is the power of the $i$-th source signal for all $t\in[T]$. The additive noises follow $\mathbf{n}(t)\sim\mathcal{CN}(\mathbf{0},\eta\mathbf{I}_M)$ for all $t\in[T]$ which are i.i.d. and uncorrelated with $\mathbf{s}(t)$ for all $t\in[T]$. We further assume that $T\geq M$.

Let $N\leq M$ and $\mathcal{S}=\{s_1,s_2,\cdots,s_N\}\subset[M]$ such that $s_1<s_2<\cdots<s_N$. Then a physical $N$-element linear array can be created by removing the $i$-th sensor from a virtual $M$-element ULA if $i\not\in\mathcal{S}$ for all $i\in[M]$. As a result, the snapshot $\mathbf{y}_{\mathcal{S}}(t)\in\mathbb{C}^N$ received on this physical $N$-element linear array at time $t\in[T]$ is given by
$
    \mathbf{y}_{\mathcal{S}}(t)=\boldsymbol{\Gamma}\mathbf{y}(t)
$
where $\mathbf{y}(t)$ is the snapshot received on the virtual $M$-element ULA and $\boldsymbol{\Gamma}\in\mathbb{R}^{N\times M}$ is a row selection matrix given by
\begin{equation}
    \left[\boldsymbol{\Gamma}\right]_{nm}=
    \begin{cases}
    1, &\text{ if } s_n=m,\\
    0, &\text{ otherwise},
    \end{cases}, n\in[N], m\in[M].
\end{equation}
In this paper, we are interested in $\mathcal{S}$ that gives rise to an SLA with the same aperture as the ULA and with no holes in its co-array such as an MRA \cite{van2002optimum} or a nested array \cite{pal2010nested}. The number of sources $k$ is assumed to be given.

\subsection{SCMs and the DoA estimation problem}
\noindent With the above assumptions, it follows that the noiseless SCM of the ULA at every $t\in[T]$ can be written as
$
    \mathbf{R}_0=\mathbf{A}(\boldsymbol{\theta})\mathbf{P}\mathbf{A}^{\mathsf{H}}(\boldsymbol{\theta})
$
and the noiseless SCM of the SLA is
$
    \mathbf{R}_{\mathcal{S}}=\boldsymbol{\Gamma}\mathbf{R}_0\boldsymbol{\Gamma}^{\mathsf{T}}.
$
The sample SCM of the ULA and the SLA are denoted by $\hat{\mathbf{R}}=\frac{1}{T}\sum_{t=1}^T\mathbf{y}(t)\mathbf{y}^{\mathsf{H}}(t)$ and
$
    \hat{\mathbf{R}}_{\mathcal{S}}=\frac{1}{T}\sum_{t=1}^T\mathbf{y}_{\mathcal{S}}(t)\mathbf{y}_{\mathcal{S}}^{\mathsf{H}}(t).
$
Given $M$, $k$, $\mathcal{S}$, and $\hat{\mathbf{R}}_{\mathcal{S}}$, the goal of the DoA estimation problem is to recover $\boldsymbol{\theta}$. Note that it is possible that $k>N$ because $k\in[M-1]$. In this paper, we focus on gridless methods recovering an $M$-by-$M$ matrix and use the root-MUSIC algorithm \cite{barabell1983improving,rao1989performance} to find $\boldsymbol{\theta}$.

\subsection{Neural network models}
\noindent The rectified linear unit (ReLU) activation function is defined as
$
    x\mapsto\max(0,x)
$. A ReLU network can be expressed as a composition of affine functions and ReLU activation functions. We adopt the definition of ReLU networks from Definition 4 in \cite{chen2022improved}. Given a complex-valued input, we first separate it into its real and imaginary components, which are then processed by the network. The network produces corresponding real-valued outputs, which are subsequently recombined to form a complex-valued result.
\section{Prior Art} \label{sec:prior_art}
\noindent According to the assumptions and settings in Section \ref{sec:prelim}, we will briefly review several popular or insightful approaches in the literature including the widely used SDP-based methods and recently proposed DNN-based approaches. Despite their differences, notice that most of them fall into the category of minimizing some distance between two covariance matrices in an appropriate space. The materials covered in this section will serve as important background and contrast with our main contributions detailed in Section \ref{sec:subspace_representation_learning}.
\subsection{The maximum likelihood problem}
\noindent Based on the assumptions in Section \ref{assumptions}, it follows that $\mathbf{R}_0+\eta\mathbf{I}_M$ is positive semidefinite and possibly Toeplitz. Because $\mathbf{y}_{\mathcal{S}}(t)\sim\mathcal{CN}(\mathbf{0},\mathbf{R}_{\mathcal{S}}+\eta\mathbf{I}_N)$, one can formulate the following constrained optimization problem according to the maximum likelihood principle:
\begin{equation} \label{prob:ml}
\begin{split}
    &\min_{\mathbf{R}\in\mathbb{H}^M} \quad \log\det\left(\boldsymbol{\Gamma}\mathbf{R}\boldsymbol{\Gamma}^{\mathsf{T}}\right)+\text{tr}\left(\left(\boldsymbol{\Gamma}\mathbf{R}\boldsymbol{\Gamma}^{\mathsf{T}}\right)^{-1}\hat{\mathbf{R}}_{\mathcal{S}}\right)\\
    &\text{subject to} \quad \mathbf{R}\succeq 0, \quad \mathbf{R}\in\mathbb{T}^M.
\end{split}
\end{equation}
By minimizing the Kullback–Leibler divergence of $\mathcal{CN}\left(\mathbf{0},\hat{\mathbf{R}}_{\mathcal{S}}\right)$ from $\mathcal{CN}\left(\mathbf{0},\boldsymbol{\Gamma}\mathbf{R}\boldsymbol{\Gamma}^{\mathsf{T}}\right)$, one can also derive the above problem. Due to the nonconvex objective, solving (\ref{prob:ml}) is nontrivial; and thus relaxing or refomulating (\ref{prob:ml}) into a tractable problem is often necessary to arrive at an accepted solution. Section \ref{subsec:redundancy_avg} and \ref{subsec:sdp} below describe tractable optimization problems that are widely used in this context.
\subsection{Redundancy averaging and direct augmentation} \label{subsec:redundancy_avg}
\noindent Because an SLA can generate all of the autocorrelation lags of the corresponding ULA, Pillai \textit{et al.} proposed the earliest approach of recovering $\mathbf{R}_0+\eta\mathbf{I}_M$ from $\hat{\mathbf{R}}_{\mathcal{S}}$, i.e., the so-called redundancy averaging and direct augmentation approach \cite{pillai1985new}. This approach is identical to solving the following matrix augmentation problem \cite{wang2019grid}:
\begin{equation} \label{prob:da}
    \min_{\mathbf{R}\in\mathbb{C}^{M\times M}} \quad \left\lVert\boldsymbol{\Gamma}\mathbf{R}\boldsymbol{\Gamma}^{\mathsf{T}}-\hat{\mathbf{R}}_{\mathcal{S}}\right\rVert_F \quad \text{subject to} \quad \mathbf{R}\in\mathbb{T}^M
\end{equation}
which has a closed-form solution that is Hermitian and Toeplitz but not necessarily positive semidefinite. Spatial smoothing \cite{pal2010nested} can be applied to fix this issue via $\frac{1}{M}\mathbf{R}\mathbf{R}^{\mathsf{H}}$ if $\mathbf{R}$ is the solution of (\ref{prob:da}).
\subsection{Direct SDP-based methods} \label{subsec:sdp}
\noindent Based on the covariance fitting criterion \cite{stoica2010new}, Yang \textit{et al.} formulated the SPA \cite{yang2014discretization} involving the optimization problem:
\begin{equation} \label{prob:spa}
\begin{split}
    &\min_{\mathbf{X}\in\mathbb{H}^N,\mathbf{R}\in\mathbb{H}^M} \quad \text{tr}\left(\mathbf{X}\right) + \text{tr}\left(\hat{\mathbf{R}}_{\mathcal{S}}^{-1}\boldsymbol{\Gamma}\mathbf{R}\boldsymbol{\Gamma}^{\mathsf{T}}\right)\\
    &\text{subject to} \quad \begin{bmatrix}\mathbf{X}&\hat{\mathbf{R}}_{\mathcal{S}}^{\frac{1}{2}}&\\
    \hat{\mathbf{R}}_{\mathcal{S}}^{\frac{1}{2}}&\boldsymbol{\Gamma}\mathbf{R}\boldsymbol{\Gamma}^{\mathsf{T}}\\
    & & \mathbf{R}\end{bmatrix}\succeq 0, \quad \mathbf{R}\in\mathbb{T}^M.
\end{split}
\end{equation}
The noiseless SCM is then estimated by $\mathbf{R}-\lambda_{\text{min}}\left({\mathbf{R}}\right)\mathbf{I}_M$ where $\mathbf{R}$ is the solution of (\ref{prob:spa}).
Another interesting approach based on the Bures-Wasserstein distance \cite{bhatia2019bures} was developed by Wang \textit{et al.} \cite{wang2019grid}. The optimization problem is given by
\begin{equation} \label{prob:wg}
\begin{split}
    &\min_{\mathbf{X}\in\mathbb{C}^{N\times N},\mathbf{R}\in\mathbb{H}^M} \quad \text{tr}\left(\hat{\mathbf{R}}_{\mathcal{S}}+\boldsymbol{\Gamma}\mathbf{R}\boldsymbol{\Gamma}^{\mathsf{T}}-\mathbf{X}-\mathbf{X}^{\mathsf{H}}\right)\\
    &\text{subject to} \quad \begin{bmatrix}\boldsymbol{\Gamma}\mathbf{R}\boldsymbol{\Gamma}^{\mathsf{T}}&\mathbf{X}\\
    \mathbf{X}^{\mathsf{H}}&\hat{\mathbf{R}}_{\mathcal{S}}\end{bmatrix}\succeq 0, \quad \mathbf{R}\succeq 0, \quad \mathbf{R}\in\mathbb{T}^M.
\end{split}
\end{equation}
Both optimization problems in (\ref{prob:spa}) and (\ref{prob:wg}) are SDPs that can be solved by off-the-shelf solvers such as the SDPT3 \cite{toh1999sdpt3}.

\subsection{Majorization-Minimization}
\noindent Since the term $\log\det(\cdot)$ in (\ref{prob:ml}) is concave on the positive semidefinite cone and the trace term can be written as an SDP via the Schur complement lemma, majorization-minimization algorithms can be used to tackle (\ref{prob:ml}). Using a supporting hyperplane to majorize the term $\log\det$, one can derive the so-called ``StructCovMLE'' approach \cite{pote2023maximum}. Let $\mathbf{R}^{(0)}$ be initialized to $\mathbf{I}_M$. For $i=0,1,2,\cdots$, StructCovMLE calculates the iterate $\mathbf{R}^{(i+1)}$ by solving the optimal $\mathbf{R}$ in the following SDP:
\begin{equation} \label{prob:structcovmle}
\begin{split}
& \min_{\mathbf{R}\in\mathbb{H}^M,\mathbf{X}\in\mathbb{H}^N} \text{tr}\left(\left(\boldsymbol{\Gamma}\mathbf{R}^{(i)}\boldsymbol{\Gamma}^{\mathsf{T}}\right)^{-1}\boldsymbol{\Gamma}\mathbf{R}\boldsymbol{\Gamma}^{\mathsf{T}}\right)+\text{tr}\left(\mathbf{X}\hat{\mathbf{R}}_{\mathcal{S}}\right)\\
&\text{subject to} \quad \begin{bmatrix}\mathbf{X}&\mathbf{I}_N&\\\mathbf{I}_N&\boldsymbol{\Gamma}\mathbf{R}\boldsymbol{\Gamma}^{\mathsf{T}}&\\ & & \mathbf{R}\end{bmatrix}\succeq 0, \quad \mathbf{R}\in\mathbb{T}^M.
\end{split}
\end{equation}
The final solution is then obtained through running a number of iterations until a stopping criterion is satisfied. For example, the relative change between $\mathbf{R}^{(i)}$ and $\mathbf{R}^{(i+1)}$ being sufficiently small. As there is a sequence of SDPs to be solved, the complexity of this approach is greater than the complexity of the above direct SDP-based methods in Section \ref{subsec:sdp}.

\subsection{Proxy covariance matrix estimation} \label{sec:prox_cov}
\noindent Instead of estimating the covariance matrix, Sarangi \textit{et al.} \cite{sarangi2021beyond} proposed a ``proxy covariance matrix'' approach (Prox-Cov) that jointly calculates a positive definite weighting matrix $\mathbf{W}$ and a proxy covariance $\mathbf{R}$ such that the weighted covariance matrix from the data best fits the proxy covariance. Based on this rationale, they formulated the following SDP:
\begin{equation} \label{prob:prox_cov}
\begin{split}
& \min_{\mathbf{R}\in\mathbb{H}^M,\mathbf{W}\in\mathbb{H}^T} \quad \left\lVert \mathbf{Y}\mathbf{W}\mathbf{Y}^{\mathsf{H}} - \boldsymbol{\Gamma}\mathbf{R}\boldsymbol{\Gamma}^{\mathsf{T}} \right\rVert_F^2\\
&\text{subject to} \quad \mathbf{R}\succeq 0, \quad \mathbf{R}\in\mathbb{T}^M, \quad \mathbf{W}\succeq \epsilon\mathbf{I}_T
\end{split}
\end{equation}
where $\mathbf{Y}=\begin{bmatrix}\mathbf{y}(1)&\mathbf{y}(2)&\cdots&\mathbf{y}(T)\end{bmatrix}$ is a matrix whose columns are all of the received snapshots and $\epsilon$ is a hyperparameter which is strictly positive. An interesting property of (\ref{prob:prox_cov}) is that it can exactly recover the signal subspace, overcoming the shortcoming of (\ref{prob:da}), under appropriate assumptions \cite{sarangi2021beyond}.
Unlike the aforementioned methods that estimate a covariance matrix from a sample SCM, Prox-Cov considers all snapshots and attempts to estimate the signal and noise subspaces by introducing a weighting matrix $\mathbf{W}$, which allows for arbitrary signal powers while maintaining the same range space from the snapshots.

\subsection{DNN-based covariance matrix reconstruction}
\noindent Let $\mathcal{D}=\left\{\hat{\mathbf{R}}_{\mathcal{S}}^{(l)},\mathbf{R}_0^{(l)}\right\}_{l=1}^L$ be a dataset containing $L$ pairs of matrices where every $\hat{\mathbf{R}}_{\mathcal{S}}^{(l)}\in\mathbb{H}^{N}$ is a sample SCM of the $N$-element SLA and $\mathbf{R}_0^{(l)}\in\mathbb{H}^M$ is the corresponding noiseless SCM of the $M$-element ULA. According to the work by Barthelme and Utschick \cite{barthelme2021doa}, one can formulate the matrix estimation problem as a learning problem whose goal is to find optimal parameters $W^*$ of a DNN model $f_W:\mathbb{C}^{N\times N}\to\mathbb{C}^{M\times M}$ such that $f_{W^*}\left(\hat{\mathbf{R}}_{\mathcal{S}}\right)f_{W^*}^{\mathsf{H}}\left(\hat{\mathbf{R}}_{\mathcal{S}}\right)\approx\mathbf{R}_0$ for every possible pair $\left(\hat{\mathbf{R}}_{\mathcal{S}},\mathbf{R}_0\right)$ of interest. The Gram matrix here is to ensure the positive semidefiniteness. The search of $W^*$ is done through the training of the DNN. After training, the function $f_{W^*}$ is evaluated at an $N$-by-$N$ sample SCM to obtain an $M$-by-$M$ SCM estimate. The model $f_W$ is trained by solving the empirical risk minimization problem
\begin{equation} \label{old:erm}
    \min_{W} \quad \frac{1}{L}\sum_{l=1}^Ld\left(f_W\left(\hat{\mathbf{R}}_{\mathcal{S}}^{(l)}\right)f_W^{\mathsf{H}}\left(\hat{\mathbf{R}}_{\mathcal{S}}^{(l)}\right),\mathbf{R}_0^{(l)}\right)
\end{equation}
where $d$ is a metric or distance. For example, the well-known Frobenius norm
\begin{equation}
    d_{\text{Fro}}\left(\mathbf{E},\mathbf{F}\right)=\left\lVert\mathbf{E}-\mathbf{F}\right\rVert_F
\end{equation}
and the \textit{affine invariant distance} \cite{bhatia2009positive}
\begin{equation} \label{eq:affine_invariant_distance}
    d_{\text{Aff}}\left(\mathbf{E},\mathbf{F}\right)=\left\lVert\log\left(\mathbf{F}^{-\frac{1}{2}}\mathbf{E}\mathbf{F}^{-\frac{1}{2}}\right)\right\rVert_F
\end{equation}
that gives the length of the shortest curve between the two points in the convex cone of all positive definite matrices $\{\mathbf{E}\in\mathbb{H}^M \mid \mathbf{E}\succ 0\}$. If (\ref{eq:affine_invariant_distance}) is used in (\ref{old:erm}), $\mathbf{R}_0^{(l)}$ is replaced by $\mathbf{R}_0^{(l)}+\delta\mathbf{I}_M$ for some $\delta>0$ as $\mathbf{R}_0^{(l)}$ can be singular.
Although this method by Barthelme and Utschick \cite{barthelme2021doa} was originally developed for the subarray sampling problem, we find that it is suitable for the matrix estimation problem in this paper.

An early study in the literature addressing the matrix estimation problem using a DNN is the work of Wu \textit{et al.} \cite{wu2022gridless}. Let $\mathbf{u}\in\mathbb{C}^M$ be the vector representing the first row of $\mathbf{A}(\boldsymbol{\theta})\mathbf{A}^{\mathsf{H}}(\boldsymbol{\theta})$. Instead of using the Gram matrix to generate a positive semidefinite matrix output, Wu \textit{et al.} constructed a DNN $f_{W_k}:\mathbb{C}^{N\times N}\to\mathbb{C}^{M}$ to estimate $\mathbf{u}$ and then recovered the matrix by $\text{Toep}\left(f_{W_k}\left(\hat{\mathbf{R}}_{\mathcal{S}}\right)\right)$ for a given $\hat{\mathbf{R}}_{\mathcal{S}}$ and source number $k$. The models $\left\{f_{W_k}\right\}_{k=1}^{M-1}$ were trained individually by the squared loss function
\begin{equation}
    d_{\text{squ}}\left(\mathbf{u},\mathbf{v}\right)=\frac{1}{2M}\left\lVert\mathbf{u}-\mathbf{v}\right\rVert_2^2.
\end{equation}
Though $M-1$ DNNs are used in \cite{wu2022gridless}, note that this method is not limited by the number of DNNs used. The Toeplitz prior and $d_{\text{squ}}$ can be used to train a single network if desired.
\section{Subspace Representation Learning} \label{sec:subspace_representation_learning}
\noindent A weakness of the above DNN-based methods is that their loss functions are not invariant to a different matrix representation of the signal or noise subspace. To elaborate, let $\boldsymbol{\Sigma}\in\mathbb{H}^{K}$ be any positive definite matrix such that $\boldsymbol{\Sigma}\neq\mathbf{P}$. Then, $\mathbf{A}(\boldsymbol{\theta})\boldsymbol{\Sigma}\mathbf{A}^{\mathsf{H}}(\boldsymbol{\theta})$ and $\mathbf{R}_0$ have exactly the same signal subspace
$
    \left\{\mathbf{A}(\boldsymbol{\theta})\mathbf{x} \middle| \mathbf{x}\in\mathbb{C}^{K}\right\}
$
that leads to the same DoAs via the root-MUSIC algorithm. However, $\mathbf{A}(\boldsymbol{\theta})\boldsymbol{\Sigma}\mathbf{A}^{\mathsf{H}}(\boldsymbol{\theta})\neq\mathbf{R}_0$ which implies $d\left(\mathbf{A}(\boldsymbol{\theta})\boldsymbol{\Sigma}\mathbf{A}^{\mathsf{H}}(\boldsymbol{\theta}),\mathbf{R}_0\right)>0$ for any metric or distance $d$ on $\mathbb{C}^{M\times M}$. If $\boldsymbol{\Sigma}=\rho\mathbf{I}_K$, it can be easily seen that $d\to\infty$ as $\rho\to\infty$ for most of the common distances such as $d_{\text{Fro}}$ and $d_{\text{squ}}$ mentioned above even though the signal subspace induced by $\mathbf{A}(\boldsymbol{\theta})\boldsymbol{\Sigma}\mathbf{A}^{\mathsf{H}}(\boldsymbol{\theta})$ is always the same as the one induced by $\mathbf{R}_0$. This is not a desirable property for a loss function because it significantly reduces the solution space and makes it much more difficult to find and approximate the signal or noise subspace.
It is worth noting that many existing methods (e.g., most of the methods in Section \ref{sec:prior_art}) measure the goodness of fit via some distance between two covariance matrices, effectively solving a harder problem than needed. Because the root-MUSIC algorithm only requires the knowledge of the signal or noise subspace, the problem of covariance estimation is actually harder than DoA estimation.

To address the above-mentioned issue, we propose a new methodology which we call \textit{subspace representation learning}. In the subsections below, we will first introduce a new output representation for DNN models to establish the invariance to the choice of $\boldsymbol{\Sigma}$. Next, we construct a novel family of loss functions to train these DNN models based on the goodness of subspace fitting and show that it is possible for a DNN to approximate signal subspaces. The root-MUSIC algorithm is then applied on the learned signal subspace to obtain the DoAs. We then discuss the use case for imperfect arrays. Finally, we propose a new batch sampling approach to parallelize the computation involved during training.

\subsection{Subspace representations of different dimensions} \label{sec:codomain_and_data}
\noindent Because every $k$-dimensional subspace $\mathcal{U}_k$ of $\mathbb{C}^M$ is a point in the \textit{Grassmann manifold} or \textit{Grassmannian} $\text{Gr}(k,M)$, we construct a DNN model $f_W$ such that
\begin{equation} \label{eq:model}
    f_W:\mathbb{C}^{N\times N}\times [M-1]\to\bigcup_{k=1}^{M-1}\text{Gr}(k,M).
\end{equation}
The codomain is a union of $M-1$ Grassmannians. To represent points of this union numerically, we can pick any matrix $\mathbf{U}\in\mathbb{C}^{M\times k}$ whose columns represent a unitary basis of $\mathcal{U}_k\in\text{Gr}(k,M)$ for all $k\in[M-1]$. Based on this perspective, the model $f_W$ is instructed to generate a matrix $\mathbf{X}\in\mathbb{C}^{M\times M}$ whose Gram matrix is factorized by eigenvalue decomposition:
\begin{equation}
\mathbf{X}\mathbf{X}^{\mathsf{H}}=\begin{bmatrix}\tilde{\mathbf{U}}&\tilde{\mathbf{V}}\end{bmatrix}\begin{bmatrix}\boldsymbol{\Lambda}_k&\\&\boldsymbol{\Lambda}_{M-k}\end{bmatrix}\begin{bmatrix}\tilde{\mathbf{U}}^{\mathsf{H}}\\\tilde{\mathbf{V}}^{\mathsf{H}}\end{bmatrix}
\end{equation}
where $\boldsymbol{\Lambda}_k$ and $\boldsymbol{\Lambda}_{M-k}$ are diagonal matrices representing the $k$ largest eigenvalues and $M-k$ smallest eigenvalues, respectively; and the columns of $\tilde{\mathbf{U}}$ and $\tilde{\mathbf{V}}$ are their corresponding orthonormal eigenvectors, respectively. Since the columns of $\tilde{\mathbf{U}}\in\mathbb{C}^{M\times k}$ form a unitary basis, a subspace $\tilde{\mathcal{U}}_k\in\text{Gr}(k,M)$ can then be identified by the range space of $\tilde{\mathbf{U}}$ and thus the function $f_W$ can generate points in the union of the Grassmannians. As long as $\mathbf{X}$ maintains the same signal subspace, the subspace $\tilde{\mathcal{U}}$ generated by $f_W$ is invariant to the change of $\mathbf{X}$. One simple invariance can be easily seen by changing the eigenvalues while maintaining the order of $\boldsymbol{\Lambda}_k$ and $\boldsymbol{\Lambda}_{M-k}$. The subspace $\tilde{\mathcal{U}}$ is also invariant to the equivalence class of its unitary bases.

Given a dataset $\mathcal{D}=\left\{\hat{\mathbf{R}}_{\mathcal{S}}^{(l)},\mathbf{R}_0^{(l)}\right\}_{l=1}^L$, we extract the signal subspace $\mathcal{U}^{(l)}$ of $\mathbf{R}_0^{(l)}$ for every $l\in[L]$ via eigenvalue decomposition to create target subspace representations. Note that $\mathcal{U}^{(l)}$ can also be identified from $\mathbf{A}\left(\boldsymbol{\theta}^{(l)}\right)\mathbf{A}^{\mathsf{H}}\left(\boldsymbol{\theta}^{(l)}\right)$ if only a dataset of $\left\{\hat{\mathbf{R}}_{\mathcal{S}}^{(l)},\boldsymbol{\theta}^{(l)}\right\}$ is available.

\subsection{Distances between subspace representations}
\noindent To learn the target subspace representations in $\mathcal{D}$, we find the parameters $W$ by solving the following empirical risk minimization problem
\begin{equation} \label{erm:subspace_representation_learning}
    \min_{W} \quad \frac{1}{L}\sum_{l=1}^Ld_{k=k^{(l)}}\left(f_W\left(\hat{\mathbf{R}}_{\mathcal{S}}^{(l)},k^{(l)}\right),\mathcal{U}^{(l)}\right)
\end{equation}
where $d_k:\text{Gr}(k,M)\times\text{Gr}(k,M)\to[0,\infty)$ is some distance on the Grassmannian $\text{Gr}(k,M)$. We propose to construct $d_k$ as a function of the vector of \textit{principal angles} between two given subspaces because it is a necessary condition if $d_k$ is invariant to any rotation in the unitary group $\mathbb{U}(M)$ of $M$-by-$M$ unitary matrices \cite{wong1967differential}, i.e.,
\begin{equation} \label{eq:rotational_invariance}
    d_k\left(\mathbf{Q}\cdot\mathcal{U},\mathbf{Q}\cdot\tilde{\mathcal{U}}\right)=d_k\left(\mathcal{U},\tilde{\mathcal{U}}\right)
\end{equation}
for every $\mathcal{U},\tilde{\mathcal{U}}\in\text{Gr}(k,M)$ and every $\mathbf{Q}\in\mathbb{U}(M)$. The \textit{left action} of $\mathbb{U}(M)$ on $\text{Gr}(k,M)$ in (\ref{eq:rotational_invariance}) is defined by
$
    \mathbf{Q}\cdot\mathcal{U}\coloneqq\text{span}\left(\mathbf{Q}\mathbf{B}\right)
$ where the columns of $\mathbf{B}\in\mathbb{C}^{M\times k}$ form a basis of $\mathcal{U}$.
According to Theorem 1 of \cite{bjorck1973numerical}, the principal angles $\boldsymbol{\phi}_k=\begin{bmatrix}\phi_1&\phi_2&\cdots&\phi_k\end{bmatrix}^{\mathsf{T}}$ between $\mathcal{U}\in\text{Gr}(k,M)$ and $\tilde{\mathcal{U}}\in\text{Gr}(k,M)$ can be calculated by
\begin{equation}
    \phi_i\left(\mathcal{U},\tilde{\mathcal{U}}\right)=\cos^{-1}\left(\sigma_i\left(\mathbf{U}^{\mathsf{H}}\tilde{\mathbf{U}}\right)\right)
\end{equation}
for $i\in[k]$ where $\mathbf{U}\in\mathbb{C}^{M\times k}$ and $\tilde{\mathbf{U}}\in\mathbb{C}^{M\times k}$ are matrices whose columns form unitary bases of $\mathcal{U}$ and $\tilde{\mathcal{U}}$, respectively, and $\sigma_1\geq\sigma_2\geq\cdots\geq\sigma_k$ are the singular values of the singular value decomposition of $\mathbf{U}^{\mathsf{H}}\tilde{\mathbf{U}}$. As $\cos^{-1}$ is a monotonically decreasing function over its domain, the principal angles satisfy $\phi_1\leq\phi_2\leq\cdots\leq\phi_k$.

Several examples of distances based on $\boldsymbol{\phi}_k$ \cite{edelman1998geometry,barg2002bounds,hamm2008grassmann,ye2016schubert} are provided in Table \ref{tab:subspace_distances}.
\begin{table}[htbp]
    \caption{Distances between subspaces}
    \centering
    \begin{tabular}{lc}
        \toprule
        Distance & Function of principal angles\\
        \midrule
        Geodesic (arc length) & $\left\lVert\boldsymbol{\phi}_k\right\rVert_2$\\
        Fubini-Study & $\cos^{-1}\left(\prod_{i=1}^k\cos\phi_i\right)$\\
        Chordal (projection Frobenius norm) & $\left(\sum_{i=1}^k\sin^2\phi_i\right)^{\frac{1}{2}}$\\
        Projection $2$-norm & $\sin\phi_k$ \\
        Chordal Frobenius norm & $2\left(\sum_{i=1}^k\sin^2\frac{\phi_i}{2}\right)^{\frac{1}{2}}$ \\
        Chordal $2$-norm & $2\sin\frac{\phi_k}{2}$ \\
        \bottomrule
    \end{tabular}
    \label{tab:subspace_distances}
\end{table}
Among them, the most natural choice of $d_k$ is the \textit{geodesic distance} \cite{wong1967differential}
\begin{equation} \label{eq:geodesic_distance}
    d_k^{\text{Geo}}\left(\mathcal{U},\tilde{\mathcal{U}}\right)=\left\lVert\boldsymbol{\phi}_k\left(\mathcal{U},\tilde{\mathcal{U}}\right)\right\rVert_2=\left(\sum_{i=1}^k\phi_i^2\left(\mathcal{U},\tilde{\mathcal{U}}\right)\right)^{\frac{1}{2}}
\end{equation}
which defines the length of the shortest curve between the two points $\mathcal{U}$ and $\tilde{\mathcal{U}}$ on the Grassmannian $\text{Gr}(k,M)$. The geodesic distance of any two points on $\text{Gr}(k,M)$ is bounded from above by
$
    \sqrt{k}\frac{\pi}{2}
$
\cite{wong1967differential}; and one can easily construct different loss functions which are bounded.

\subsection{Approximation}
\noindent In this subsection, we attempt to enhance the feasibility of subspace representation learning from an approximation viewpoint. In particular, we present a guarantee for a neural network model to approximate the signal subspace.
\begin{theorem} \label{thm:subspace_approximation}
    For every $k\in[M-1]$ and every $\epsilon>0$, there exists a ReLU network $f:\mathbb{C}^{N\times N}\to\text{Gr}(k,M)$ such that
    \begin{equation}
        \int_{[0,\pi]^k}d_k^{\text{Geo}}\left( f\left(\mathbf{R}_{\mathcal{S}}\right),P_{\mathbf{A}(\boldsymbol{\theta})}\right)d\boldsymbol{\theta}<\epsilon.
    \end{equation}
\end{theorem}
The proof of Theorem \ref{thm:subspace_approximation} is contained in Appendix \ref{proof:subspace_approximation}.
Here, subspaces are represented by their orthogonal projectors to ensure every $\mathcal{U}\in\text{Gr}(k,M)$ has a unique representation. In other words, $\text{Gr}(k,M)$ is equivalent to
\begin{equation}
    \left\{ P\in\mathbb{C}^{M\times M} \mid P^{\mathsf{H}}=P, P^2=P, \text{rank}(P)=k \right\}.
\end{equation}
If the ideal covariance matrices are used, Theorem \ref{thm:subspace_approximation} shows that the average geodesic distance between the predicted subspaces and the desirable signal subspaces can be made arbitrarily small when a suitable ReLU network is picked. From an array processing point of view, it is trivial that the signal subspace can always be extracted from $\mathbf{R}_{\mathcal{S}}$. However, Theorem \ref{thm:subspace_approximation} illustrates that this process can be achieved up to a small error by evaluating a continuous piecewise linear function \cite{chen2022improved}. In order to sketch the proof, notice that a simple distance on $\text{Gr}(k,M)$ can be constructed by
\begin{equation} \label{eq:simple_distance}
    \left(\mathcal{U}_1,\mathcal{U}_2\right)\mapsto\left\lVert P_{\mathcal{U}_1}-P_{\mathcal{U}_2} \right\rVert_F.
\end{equation}
Lemma \ref{lemma:geodesic_upper_bound} below shows that the geodesic distance can be bounded from above by the composition of a strictly increasing function and the simple distance in (\ref{eq:simple_distance}), allowing us to leverage the continuity of the orthogonal projection operator in an appropriate manner to prove Theorem \ref{thm:subspace_approximation}. It may be possible to extend Theorem \ref{thm:subspace_approximation} to a more realistic case using $\hat{\mathbf{R}}_{\mathcal{S}}$ with a probabilistic guarantee.
\begin{lemma} \label{lemma:geodesic_upper_bound}
    For every $\mathcal{U}_1,\mathcal{U}_2\in\text{Gr}(k,M)$ where $k\in[M-1]$,
    \begin{equation}
        d_k^{\text{Geo}}\left(\mathcal{U}_1,\mathcal{U}_2\right)\leq\sqrt{k}\sin^{-1}\left(\frac{\left\lVert P_{\mathcal{U}_1}-P_{\mathcal{U}_2} \right\rVert_F}{\sqrt{2}}\right).
    \end{equation}
\end{lemma}
The proof of Lemma \ref{lemma:geodesic_upper_bound} is contained in Appendix \ref{proof:geodesic_upper_bound}. Lemma \ref{lemma:geodesic_upper_bound} ensures that $\left\lVert P_{\mathcal{U}_1}-P_{\mathcal{U}_2} \right\rVert_F\to 0$ implies $d_k^{\text{Geo}}\left(\mathcal{U}_1,\mathcal{U}_2\right)\to 0$.

\subsection{Learning with imperfect arrays}
\noindent Sensor arrays are not perfect in reality. For example, the array manifold may be corrupted by several imperfections including the gain bias, phase bias, sensor position error, and the intersensor mutual coupling \cite{liu2018direction}. Because model-based methods such as SDP-based approaches in (\ref{prob:spa}) and (\ref{prob:wg}) often rely on prior knowledge of the sensor positions $\mathcal{S}$ to create $\boldsymbol{\Gamma}$ in their optimization problems, they are not robust to sensor position errors; and fixing such a model mismatch is nontrivial. In contrast, our methodology does not suffer from this model mismatch issue due to its geometry-agnostic or imperfection-agnostic nature. The empirical risk minimization problem we solve in the imperfect array case is still (\ref{erm:subspace_representation_learning}). As described in the last paragraph of Section \ref{sec:codomain_and_data}, $\mathcal{U}^{(l)}$ can be identified from the ground truth $\boldsymbol{\theta}^{(l)}$; and $\hat{\mathbf{R}}_{\mathcal{S}}$ is the sample SCM from the imperfect array. Hence, both the problem formulation in (\ref{erm:subspace_representation_learning}) and the model (\ref{eq:model}) do not depend on the sensor positions. In addition, our method does not need to know the array is imperfect and the degree of imperfections. The information is already embedded in the dataset and solving (\ref{erm:subspace_representation_learning}) will enforce the DNN model to learn the subspace representations of a perfect virtual ULA from the imperfect array.

\subsection{Consistent rank sampling} \label{sec:consistent_rank_sampling}
\noindent To learn subspaces of different dimensions in one DNN model, the empirical risk minimization problem (\ref{erm:subspace_representation_learning}) requires $M-1$ loss functions $d_1,d_2,\cdots,d_{M-1}$ that calculate unitary bases of different dimensions from $1$ to $M-1$. Although (\ref{erm:subspace_representation_learning}) can be solved by the well-known minibatch stochastic gradient descent (SGD) algorithm, it is hard for the computation of different dimensions to be parallelized on a graphics processing unit (GPU). To fix this issue, we propose \textit{consistent rank sampling}, a new batch sampling strategy for learning subspaces of different dimensions in one DNN model. Instead of randomly sampling from $\mathcal{D}$, we propose randomly sampling a batch of data points whose source number $k$ is consistent from $\mathcal{D}$. This way, only one $d_k$ needs to be evaluated in every gradient step, streamlining the computation of unitary bases in the same dimension $k$. It is important to note that consistent rank sampling is a crucial strategy to make training efficient. Without this strategy, training DNNs on large datasets becomes extremely difficult due to the slow training speed. Although the strategy is developed for subspace representation learning, it is generally applicable to empirical risk minimization problems that involve loss functions of different dimensions.

\section{A Gridless End-to-end Approach} \label{sec:gridless_end2end_training}
\noindent The subspace representation learning approach utilizes the root-MUSIC algorithm on the obtained subspaces to estimate the DoAs. A natural question to study here is the following: \textit{Is it possible to bypass the root-MUSIC algorithm and directly learn a model to output the DoAs in a gridless manner?} The best-known end-to-end approach is probably the work \cite{papageorgiou2021deep} by Papageorgiou \textit{et al.}; however, it relies on a grid. The approach of mean cyclic error (MCE) network or MCENet \cite{barthelme2021machine} by Barthelme and Utschick is a gridless end-to-end approach but it was designed for subarray sampling and not for more sources than sensors. Below, we propose a new gridless end-to-end approach that is tailored to the localization of more sources than sensors using an SLA.
\usetikzlibrary{shapes,arrows}
\usetikzlibrary{positioning}
\tikzstyle{blockSo} = [draw, fill=olive!30, rectangle, minimum height=6em, minimum width=20em, line width=0.2em]
\tikzstyle{blockSb1} = [draw, fill=blue!20, rectangle, rectangle, trapezium, trapezium angle=150, minimum height=1em, line width=0.2em]
\tikzstyle{blockSb2} = [draw, fill=blue!20, rectangle, rectangle, trapezium, trapezium angle=120, minimum height=3em, line width=0.2em]
\tikzstyle{blockSb3} = [draw, fill=blue!20, rectangle, trapezium, trapezium angle=110, minimum height=4em, line width=0.2em]
\tikzstyle{blockSb4} = [draw, fill=blue!20, trapezium, trapezium angle=100, minimum height=5em, line width=0.2em]
\begin{figure}[t]
    \begin{center}
    \resizebox{0.35\textheight}{!}{
        \begin{tikzpicture}
            \node (origin) [] at (0,0) {\Huge $\hat{\mathbf{R}}_{\mathcal{S}}$};
        
            \node (backbone) [blockSo, below=1.5cm of origin] {\Huge Architecture};
            
            \node (conv3) [blockSb3, below=2cm of backbone] {\Huge Affine};
        
            \node (conv2) [blockSb2, left=2cm of conv3] {\Huge Affine};
        
            \node (conv1) [blockSb1, left=2cm of conv2] {\Huge Affine};
        
            \node (conv4) [right=2cm of conv3] {\Huge $\cdots$};
        
            \node (conv5) [blockSb4, right=2cm of conv4] {\Huge Affine};
        
            \node (out1) [below=3cm of conv1] {\Huge $\theta_1$};
            \node (out2) [below=2.7cm of conv2] {\Huge $\theta_1,\theta_2$};
            \node (out3) [below=2.5cm of conv3] {\Huge $\theta_1,\theta_2,\theta_3$};
            \node (out5) [below=2.2cm of conv5] {\Huge $\theta_1,\theta_2,\cdots,\theta_{M-1}$};
        
            \draw [ultra thick,->] (origin) -- (backbone);
        
            \draw [ultra thick,->] (backbone.south) -- (conv1.north);
            \draw [ultra thick,->] (backbone.south) -- (conv2.north);
            \draw [ultra thick,->] (backbone.south) -- (conv3.north);
            \draw [ultra thick,->] (backbone.south) -- (conv5.north);
            
            \draw [ultra thick,->] (conv1) -- (out1);
            \draw [ultra thick,->] (conv2) -- (out2);
            \draw [ultra thick,->] (conv3) -- (out3);
            \draw [ultra thick,->] (conv5) -- (out5);
        \end{tikzpicture}
}
\end{center}
\caption{An illustration of the gridless end-to-end model, which consists of an architecture and several output layers. The model simultaneously generates DoAs for every possible number of sources so there are $M-1$ heads (affine functions) at the output. The $k$-th head is picked when there are $k$ sources.}
\label{fig:gridless_end2end_architecture}
\end{figure}
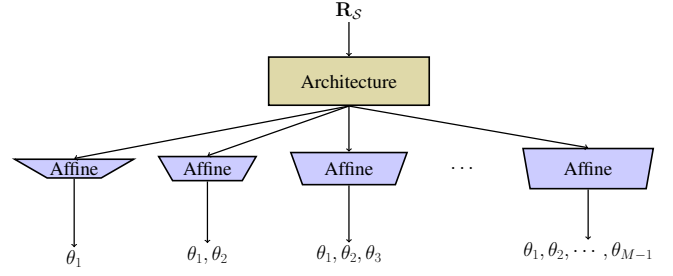
As illustrated in Fig. \ref{fig:gridless_end2end_architecture}, we propose to construct a DNN model $g_W$ such that
\begin{equation}
    g_W:\mathbb{C}^{N\times N}\times [M-1]\to\mathbb{R}^{1}\times\mathbb{R}^{2}\times\cdots\mathbb{R}^{M-1}.
\end{equation}
The codomain is the $(M-1)$-ary Cartesian product of Euclidean spaces $\mathbb{R}^1,\mathbb{R}^2,\cdots,\mathbb{R}^{M-1}$. These Euclidean spaces are viewed as different ``heads'' at the output of the model where the $k$-dimensional Euclidean space represents the $k$-th head. The $k$-th head will be picked when there are $k$ sources such that an element from $\mathbb{R}^k$ can represent $k$ angles. Let $r(i)=\frac{i(i-1)}{2}$ for $i=1,2,\cdots,M$ and denote $h_k:\mathbb{R}^{r(M)}\to\mathbb{R}^k$ the projection
\begin{equation}
\left(x_1,\cdots,x_{r(M)}\right)\mapsto \left(x_{r(k)+1},x_{r(k)+2},\cdots,x_{r(k)+k}\right).
\end{equation}
The empirical risk minimization problem of the gridless end-to-end model $g_W$ is then formulated as follows
\begin{equation} \label{erm:gridless_end2end}
    \min_{W} \quad \frac{1}{L}\sum_{l=1}^Ld_{k=k^{(l)}}\left(h_{k=k^{(l)}}\circ g_W\left(\hat{\mathbf{R}}_{\mathcal{S}}^{(l)},k^{(l)}\right),\boldsymbol{\theta}^{(l)}\right)
\end{equation}
where $d_1,d_2,\cdots,d_{M-1}$ are loss functions of different dimensions that calculate some minimum distances among all permutations. Taking the squared loss for example,
\begin{equation} \label{eq:squared_loss_optimal_permutation}
    d_k\left(\hat{\boldsymbol{\theta}},\boldsymbol{\theta}\right)=\frac{1}{k}\min_{\boldsymbol{\Pi}\in\mathcal{P}_k}\left\lVert \boldsymbol{\Pi}\hat{\boldsymbol{\theta}} - \boldsymbol{\theta} \right\rVert_2^2
\end{equation}
for $k=1,2,\cdots,M-1$. The minimum in (\ref{eq:squared_loss_optimal_permutation}) is equivalent to the squared loss applied to the corresponding sorted arguments according to the rearrangement inequality \cite{hardy1934inequalities}. Because a loss function of different dimensions is adopted, the consistent rank sampling strategy detailed in Section \ref{sec:consistent_rank_sampling} can be applied to accelerate training on GPUs.
\section{Numerical Results} \label{sec:numerical_results}
\noindent In this section, we will compare our new methodologies with existing methods including the SPA \cite{yang2014discretization}, the Wasserstein distance based approach (WDA) \cite{wang2019grid}, the DNN-based covariance reconstruction (DCR) approach based on the Toeplitz prior \cite{wu2022gridless} (DCR-T), and the DCR approach based on the Gram matrix \cite{barthelme2021doa} (DCR-G). In particular, we use both the Frobenius norm and the affine invariant distance for DCR-G, leading to two methods termed DCR-G-Fro and DCR-G-Aff. We do not include StructCovMLE \cite{pote2023maximum} and Prox-Cov \cite{sarangi2021beyond} because the performance of StructCovMLE was similar to SPA, and Prox-Cov \cite{sarangi2021beyond} did not yield better performance than the SPA in our preliminary experiments. In subspace representation learning, the geodesic distance in (\ref{eq:geodesic_distance}) is picked for $d_k$, if not explicitly specified.

Below, we will first set up the scenarios for the DoA estimation problem. Next, we will describe the DNN architectures and the training procedures for the DNN-based approaches. Finally, for a given SNR and number of snapshots $T$, we will compare performance of different approaches in terms of the mean squared error (MSE)
\begin{equation}
    \frac{1}{L_{\text{test}}}\sum_{l=1}^{L_{\text{test}}}\frac{1}{k}\min_{\boldsymbol{\Pi}\in\mathcal{P}_k}\left\lVert\boldsymbol{\Pi}\hat{\boldsymbol{\theta}}_{l}-\boldsymbol{\theta}_{l}\right\rVert_2^2
\end{equation}
for different source numbers $k\in[M-1]$ where $L_{\text{test}}$ is the total number of random trials, $\boldsymbol{\theta}_{l}$ is the vector of DoAs of the ground truth at the $l$-th trial, and $\hat{\boldsymbol{\theta}}_{l}$ is the corresponding estimate given by a method of interest. Code is available at
{\fontfamily{qcr}\selectfont
\url{https://github.com/kjason/SubspaceRepresentationLearning}}.

\subsection{Settings} \label{sec:settings}
\noindent The physical array is an $N$-element MRA with $N=5$ and $\mathcal{S}=\{1,2,5,8,10\}$, leading to a $10$-element virtual ULA or $M=10$. A study for different MRAs is deferred to Section \ref{sec:4_and_6_mra}. Below we describe the test or evaluation conditions. The number of snapshots $T$ is set to $50$, if not explicitly specified. The SNR is defined as $10\log_{10}\left(\frac{\frac{1}{k}\sum_{i=1}^kp_i}{\eta}\right)$ and we assume equal source powers $p_1=p_2=\cdots=p_k$, if not explicitly specified. The SNR is set to $20$ dB if not explicitly stated. The finite set of SNRs $\{-10,-8,-6,\cdots,16,18,20\}$ is picked when a range of SNRs is required for evaluation. The number of sources $k$ can be any number in the set $[M-1]$. For any $k\in[M-1]$, the DoAs $\theta_1,\theta_2,\cdots,\theta_k$ are uniformly selected at random in the range $\left[\frac{1}{6}\pi,\frac{5}{6}\pi\right]$ with a minimum separation constraint $\min_{i\neq j}\lvert\theta_i-\theta_j\rvert\geq \frac{1}{45}\pi$. For every given SNR, $T$, and $k\in[M-1]$, there are $100$ trials of random source signals and noises for a given $\boldsymbol{\theta}$, and there are in total $100$ random $\boldsymbol{\theta}$, leading to a total number of trials $L_{\text{test}}=10^4$ for each case. All SDP problems are solved by the SDPT3 \cite{toh1999sdpt3} solver in CVX \cite{cvx,gb08}.

\subsubsection{DNN models} \label{sec:DNN_models}
\usetikzlibrary{shapes,arrows}
\usetikzlibrary{positioning}

\tikzstyle{blockSg} = [fill=green!40!gray, rectangle,
    minimum height=3em, minimum width=3em]

\tikzstyle{blockSgb} = [draw,fill=green!40!gray, rectangle, minimum height=3em, minimum width=8em, line width=0.2em]

\tikzstyle{blockSy} = [draw, fill=yellow!20, rectangle,
    minimum height=3em, minimum width=8em, line width=0.2em]

\tikzstyle{blockSlg} = [draw, fill=gray!20, rectangle,
    minimum height=3em, minimum width=8em, line width=0.2em]

\tikzstyle{blockSb} = [draw, fill=blue!20, rectangle,
    minimum height=3em, minimum width=8em, line width=0.2em]

\tikzstyle{blockSp} = [draw, fill=pink!20, rectangle,
    minimum height=3em, minimum width=8em, line width=0.2em]

\tikzstyle{blockM} = [draw, green!40!gray, dashed, fill=white!20, rectangle, minimum height=26em, minimum width=15em, line width=0.2em]

\tikzstyle{resblockM} = [draw, black!40!gray, dashed, fill=white!20, rectangle, minimum height=11.5em, minimum width=14em, line width=0.2em]

\pgfdeclarelayer{bg}
\pgfsetlayers{bg,main}

\begin{figure}[htb]
\begin{center}
\resizebox{0.35\textheight}{!}{
    \begin{tikzpicture}

        \node (origin) [] at (0,0) {};

        \node (relu1) [blockSlg, below=1.5cm of origin] {\Huge ReLU};

        \node (conv1) [blockSy, below=1cm of relu1] {\Huge Conv};

        \node (relu2) [blockSlg, below=1cm of conv1] {\Huge ReLU};

        \node (conv2) [blockSy, below=1cm of relu2] {\Huge Conv};

        \node (out) [below=2cm of conv2] {};

        \begin{pgfonlayer}{bg}
            \node (a_block) [blockM] at (-1,-5.3){};
        \end{pgfonlayer}

        \node (a_block_name) [blockSg] at (-2.6,-1.2) {\Huge Block};

        \draw [ultra thick,->] (origin) -- (relu1);

        \draw [ultra thick,->] (relu1) -- (conv1);
        \draw [ultra thick,->] (conv1) -- (relu2);
        \draw [ultra thick,->] (relu2) -- (conv2);
        \draw [ultra thick,->] (conv2) -- (out);

        \node (SCM) [] at (7,2) {\Huge Input};

        \node (expansion) [blockSy, below=1cm of SCM] {\Huge Conv};
        \node (block1) [blockSgb, below=1.2cm of expansion] {\Huge Block};
        \node (plus1) [below=0.75cm of block1] {\Huge $+$};
        \node (x1_2) [right=0.5cm of block1] {};
        \node (x1_2_p) [above=0.2cm of x1_2] {};
        \draw [ultra thick,->] (expansion) -- node[name = x1][below,xshift=-0.15cm,yshift=0.15cm] {} (block1);
        \draw [ultra thick,-] (x1) -| (x1_2);
        \draw [ultra thick,->] (x1_2_p) |- (plus1);
        \draw [ultra thick,->] (block1) -- (plus1);

        \node (block2) [blockSgb, below=1.5cm of plus1] {\Huge Block};

        \node (proj1) [blockSy, right=0.5cm of block2] {\Huge Conv};

        \node (plus2) [below=0.75cm of block2] {\Huge $+$};
        \node (x2_2) [right=0.5cm of block2] {};
        \node (x2_2_p) [above=0.2cm of x2_2] {};
        \draw [ultra thick,->] (plus1) -- node[name = x2][below,xshift=-0.15cm,yshift=-0.05cm] {} (block2);

        \draw [ultra thick,->] (x2) -| (proj1);
        \draw [ultra thick,->] (proj1) |- (plus2);

        \draw [ultra thick,->] (block2) -- (plus2);

        \node (block3) [blockSgb, below=1.2cm of plus2] {\Huge Block};
        \node (plus3) [below=0.75cm of block3] {\Huge $+$};
        \node (x3_2) [right=0.5cm of block3] {};
        \node (x3_2_p) [above=0.2cm of x3_2] {};
        \draw [ultra thick,->] (plus2) -- node[name = x3][below,xshift=-0.15cm,yshift=0.15cm] {} (block3);
        \draw [ultra thick,-] (x3) -| (x3_2);
        \draw [ultra thick,->] (x3_2_p) |- (plus3);
        \draw [ultra thick,->] (block3) -- (plus3);

        \node (corner1) [below=1cm of plus3] {};
        \node (corner2) [right=6.5cm of corner1] {};
        \node (corner3) [above=15cm of corner2] {};

        \node (block4) [blockSgb, right=8cm of expansion] {\Huge Block};

        \draw[ultra thick,->] (plus3) -- ++(0,-1.5cm) -| ++(corner3) -| (block4);

        \node (x4) [above=0.5cm of block4] {};
        \node (proj2) [blockSy, right=0.5cm of block4] {\Huge Conv};

        \node (plus4) [below=0.75cm of block4] {\Huge $+$};
        \node (x4_2) [right=0.5cm of block4] {};
        \node (x4_2_p) [above=0.2cm of x4_2] {};

        \draw [ultra thick,->] (x4.center) -| (proj2);
        \draw [ultra thick,->] (proj2) |- (plus4);

        \draw [ultra thick,->] (block4) -- (plus4);

        \node (block5) [blockSgb, below=1.2cm of plus4] {\Huge Block};
        \node (plus5) [below=0.75cm of block5] {\Huge $+$};
        \node (x5_2) [right=0.5cm of block5] {};
        \node (x5_2_p) [above=0.2cm of x5_2] {};
        \draw [ultra thick,->] (plus4) -- node[name = x5][below,xshift=-0.15cm,yshift=0.15cm] {} (block5);
        \draw [ultra thick,-] (x5) -| (x5_2);
        \draw [ultra thick,->] (x5_2_p) |- (plus5);
        \draw [ultra thick,->] (block5) -- (plus5);

        \node (final_relu) [blockSlg, below=1.2cm of plus5] {\Huge ReLU};

        \node (avg_pool) [blockSp, below=0.8cm of final_relu] {\Huge AvgPool};

        \node (final_affine) [blockSb, below=0.8cm of avg_pool] {\Huge Affine};

        \node (X) [below=0.8cm of final_affine] {\Huge Output};

        \draw [ultra thick,->] (SCM) -- (expansion);

        \draw [ultra thick,->] (plus5) -- (final_relu);
        \draw [ultra thick,->] (final_relu) -- (avg_pool);
        \draw [ultra thick,->] (avg_pool) -- (final_affine);
        \draw [ultra thick,->] (final_affine) -- (X);

        \begin{pgfonlayer}{bg}
            \node (a_residual_block) [resblockM,below=0.3cm of expansion] {};
            \node (a_residual_block_repeat) [right=0.1cm of a_residual_block] {\Huge $\times L$};
        \end{pgfonlayer}

        \begin{pgfonlayer}{bg}
            \node (a_residual_block2) [resblockM,below=-2cm of block3] {};
            \node (a_residual_block_repeat2) [right=0.1cm of a_residual_block2] {\Huge $\times (L-1)$};
        \end{pgfonlayer}

        \begin{pgfonlayer}{bg}
            \node (a_residual_block3) [resblockM,below=-2cm of block5] {};
            \node (a_residual_block_repeat3) [right=0.1cm of a_residual_block3] {\Huge $\times (L-1)$};
        \end{pgfonlayer}
    \end{tikzpicture}
    }
\end{center}
\caption{An illustration of a $3$-stage $L$-block ResNet model \cite{he2016deep}. In the wide ResNet 16-8 (WRN-16-8) \cite{zagoruyko2016wide}, there are $L=2$ blocks per stage, leading to $16$ layers in total. The widening factor is $8$, meaning that WRN-16-8 is $8$ times wider than the original ResNet. See Section \ref{sec:DNN_models} for more details.}
\label{fig:architecture}
\end{figure}
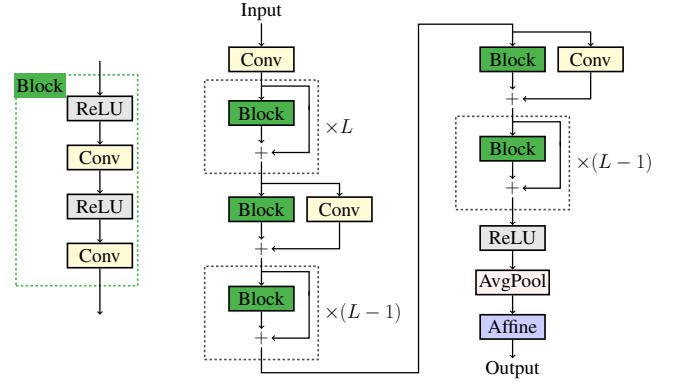
As illustrated in Fig. \ref{fig:architecture}, we use WRN-16-8 \cite{zagoruyko2016wide} without the batch normalization. The pair of numbers 16-8 implies that the total number of layers is $16$ and the widening factor is $8$. The ReLU activation function is adopted by all of the nonlinearities in the network. All of the residual blocks are in the pre-activation form \cite{he2016identity}. Note that wide ResNets avoid the degradation problem and enjoy certain optimization guarantees under mild assumptions \cite{chen2021resnests}. The network takes an input tensor in $\mathbb{R}^{2\times N\times N}$ and generates an output tensor in $\mathbb{R}^{2\times M\times M}$ ($\mathbb{R}^{2\times M}$ for DCR-T). Given an $N$-by-$N$ complex matrix, it is represented by its real and imaginary parts as inputs to the network. The first and second planes of the output tensor represent the real and imaginary parts of an $M$-by-$M$ complex matrix, respectively. The number of parameters is approximately $11$ million. All DNN-based methods use the same architecture. The output layer is an affine function whose output dimension is tailored to each approach.

\subsubsection{Training} \label{sec:training}
The minibatch SGD algorithm with Nesterov momentum is used to train all of the DNN models. The momentum is set to $0.5$ and the batch size is $4096$. The weight decay is set to $0$. All of the models are trained for $50$ epochs with the one-cycle learning rate scheduler \cite{smith2019super}. The best maximum learning rate of the scheduler for each approach is found through a grid search whose description is deferred to Appendix \ref{sec:learning_rates_grid_search}. The learning rates for DCR-T, DCR-G-Fro, DCR-G-Aff, and our approach are $0.05$, $0.01$, $0.005$, and $0.1$, respectively. The weights in all models are initialized using normal distributions \cite{he2015delving}. The value of $\delta$ is set to $10^{-4}$ in the DCR-G-Aff approach.
For each $k\in[M-1]$, there are $2\times 10^6$ and $6\times 10^5$ random data points for training and validation, respectively, leading to a training dataset of size $L_{\text{train}}=9\times 2\times 10^6$ and a validation dataset of size $L_{\text{val}}=9\times 6\times 10^5$. For each data point, the source signals and noises are generated randomly according to the assumptions in Section \ref{assumptions}. The SNR in decibels is uniformly picked at random in the finite set $\{-11,-9,-7,\cdots,17,19,21\}$. The DoAs in the vector $\boldsymbol{\theta}$ are uniformly selected at random in the range $\left[\frac{1}{6}\pi,\frac{5}{6}\pi\right]$ with a minimum separation constraint $\min_{i\neq j}\lvert\theta_i-\theta_j\rvert\geq \frac{1}{60}\pi$. The sources are assumed to have equal power, if not explicitly specified. The number of snapshots is set to $50$. PyTorch is used to train all the DNN models \cite{paszke2019pytorch}.

\subsection{Results}
\input{mse_snr_n5}
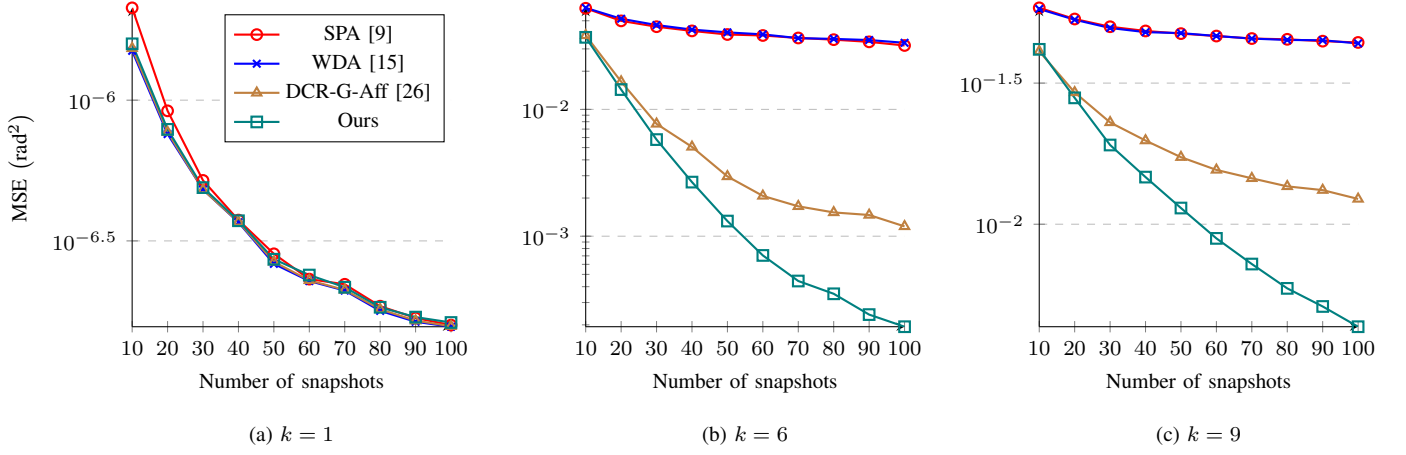
\begin{figure*}[ht]
\centering
\begin{tikzpicture}
\begin{axis}[
    font=\footnotesize,
    width=5.8cm,
    height=5.8cm,
    axis lines = left,
    xlabel = Number of snapshots,
    ylabel = MSE $\left(\text{rad}^2\right)$,
    legend style={anchor=north east},
    xtick={10,20,30,40,50,60,70,80,90,100},
    title={\footnotesize (a) $k=1$},
    title style={at={(0.5,-0.45)},anchor=south},
    xmax = 100,
    ymajorgrids=true,
    grid style=dashed,
    ymode=log
]

\addplot [
    color=red,
    mark=o,
    thick
]
coordinates {
(10,2.1279e-06)
(20,9.164e-07)
(30,5.187e-07)
(40,3.7495e-07)
(50,2.8455e-07)
(60,2.3134e-07)
(70,2.2176e-07)
(80,1.8538e-07)
(90,1.6835e-07)
(100,1.5881e-07)
};

\addlegendentry{SPA \cite{yang2014discretization}}

\addplot [
    color=blue,
    mark=x,
    thick
]
coordinates {
(10,1.4971e-06)
(20,7.5789e-07)
(30,4.8125e-07)
(40,3.6816e-07)
(50,2.6211e-07)
(60,2.2768e-07)
(70,2.1064e-07)
(80,1.7845e-07)
(90,1.6321e-07)
(100,1.5671e-07)
};

\addlegendentry{WDA \cite{wang2019grid}}

\addplot [
    color=brown,
    mark=triangle,
    thick
]
coordinates {
(10,1.5281e-06)
(20,7.7014e-07)
(30,4.8071e-07)
(40,3.6929e-07)
(50,2.6643e-07)
(60,2.2904e-07)
(70,2.1174e-07)
(80,1.808e-07)
(90,1.6469e-07)
(100,1.5738e-07)
};

\addlegendentry{DCR-G-Aff \cite{barthelme2021doa}}

\addplot [
    color=teal,
    mark=square,
    thick
]
coordinates {
(10,1.5855e-06)
(20,7.8761e-07)
(30,4.8911e-07)
(40,3.7297e-07)
(50,2.7219e-07)
(60,2.3915e-07)
(70,2.1648e-07)
(80,1.8376e-07)
(90,1.6955e-07)
(100,1.6224e-07)
};

\addlegendentry{Ours}
\end{axis}

\begin{axis}[
    font=\footnotesize,
    width=5.8cm,
    height=5.8cm,
    at={(6cm,0cm)},
    axis lines = left,
    xlabel = Number of snapshots,
    xmax = 100,
    xtick={10,20,30,40,50,60,70,80,90,100},
    title={\footnotesize (b) $k=6$},
    title style={at={(0.5,-0.45)},anchor=south},
    ymajorgrids=true,
    grid style=dashed,
    ymode=log
]

\addplot [
    color=red,
    mark=o,
    thick
]
coordinates {
(10,0.062732)
(20,0.049889)
(30,0.044887)
(40,0.041591)
(50,0.038966)
(60,0.038277)
(70,0.036577)
(80,0.035475)
(90,0.034093)
(100,0.031923)
};

\addplot [
    color=blue,
    mark=x,
    thick
]
coordinates {
(10,0.063214)
(20,0.051867)
(30,0.046253)
(40,0.042591)
(50,0.040492)
(60,0.039055)
(70,0.036471)
(80,0.035977)
(90,0.035232)
(100,0.033511)
};

\addplot [
    color=brown,
    mark=triangle,
    thick
]
coordinates {
(10,0.038348)
(20,0.016565)
(30,0.0076896)
(40,0.0050619)
(50,0.0029602)
(60,0.0020794)
(70,0.001721)
(80,0.0015401)
(90,0.0014704)
(100,0.0011951)
};

\addplot [
    color=teal,
    mark=square,
    thick
]
coordinates {
(10,0.037014)
(20,0.014337)
(30,0.0057813)
(40,0.0026713)
(50,0.0013169)
(60,0.00070543)
(70,0.00044336)
(80,0.00035157)
(90,0.00024097)
(100,0.00019338)
};

\end{axis}

\begin{axis}[
    font=\footnotesize,
    width=5.8cm,
    height=5.8cm,
    at={(12cm,0cm)},
    axis lines = left,
    xlabel = Number of snapshots,
    xmax = 100,
    xtick={10,20,30,40,50,60,70,80,90,100},
    title={\footnotesize (c) $k=9$},
    title style={at={(0.5,-0.45)},anchor=south},
    ymajorgrids=true,
    grid style=dashed,
    ymode=log
]

\addplot [
    color=red,
    mark=o,
    thick
]
coordinates {
(10,0.058438)
(20,0.053345)
(30,0.050085)
(40,0.048388)
(50,0.047352)
(60,0.046424)
(70,0.045488)
(80,0.045202)
(90,0.044525)
(100,0.044058)
};

\addplot [
    color=blue,
    mark=x,
    thick
]
coordinates {
(10,0.05789)
(20,0.053035)
(30,0.049571)
(40,0.047923)
(50,0.047545)
(60,0.046444)
(70,0.045466)
(80,0.044966)
(90,0.044857)
(100,0.043717)
};

\addplot [
    color=brown,
    mark=triangle,
    thick
]
coordinates {
(10,0.040988)
(20,0.029307)
(30,0.022947)
(40,0.019794)
(50,0.017281)
(60,0.015576)
(70,0.014543)
(80,0.013618)
(90,0.013208)
(100,0.012291)
};

\addplot [
    color=teal,
    mark=square,
    thick
]
coordinates {
(10,0.04165)
(20,0.028037)
(30,0.019085)
(40,0.014689)
(50,0.011412)
(60,0.0089094)
(70,0.007227)
(80,0.0059258)
(90,0.0051121)
(100,0.0043338)
};

\end{axis}

\end{tikzpicture}
\caption{MSE vs. number of snapshots. Although the DNN models are only trained on a single number of snapshots $T=50$, they are capable of performing well on a wide range of unseen scenarios from $T=10$ to $T=100$. Our approach is consistently better than SPA, WDA, and DCR-G-Aff.}
\label{fig:mse_vs_snapshots_n5}
\end{figure*}
\input{mse_snr_n4}
\subsubsection{Superior performance over a wide range of SNRs} \label{sec:superior_performance_SNRs}
Fig. \ref{fig:mse_vs_snr_n5} compares the proposed method with the five baseline approaches in terms of MSE over a wide range of SNRs and number of sources. For $k=1$, all of the methods have almost the same performance. For $k=2$, the proposed method is significantly better than SPA and WDA from $-10$ to $6$ dB SNR. In fact, it is uniformly better than WDA from $-10$ to $20$ dB SNR. However, once the SNR goes beyond $14$ dB, SPA starts to outperform the proposed method and the gap seems to become larger as the SNR increases. DCR-T is slightly worse than the proposed method in the high SNR region but the gap of MSE gets larger as SNR increases. With regard to DCR-G-Fro and DCR-G-Aff, their performance is similar to the proposed method. For $k=3$, the proposed method is better than SPA, WDA, DCR-T, and DCR-G-Fro across almost the entire evaluation range and even superior by orders of magnitude from $0$ to $15$ dB SNR with respect to SPA, WDA, and DCR-T. DCR-G-Aff is slightly better than the proposed method in the high SNR region but is slightly worse in the low SNR region. Then, for $k\in\{4,5,6,7,8,9\}$, the proposed method consistently and significantly outperforms SPA and WDA. As for DCR-T and DCR-G-Fro, they are noticeably better than SPA and WDA but significantly inferior than the proposed method.
In particular, DCR-G-Aff is the most competitive approach to the proposed method. However, it is still much inferior than our approach.
Overall, the proposed method is significantly better than all of the baseline approaches.

\input{mse_snr_n6}
\subsubsection{Performance on unseen numbers of snapshots}
Fig. \ref{fig:mse_vs_snapshots_n5} evaluates SPA, WDA, DCR-G-Aff, and the proposed method in terms of MSE in a wide range of numbers of snapshots and sources. We do not include the other baselines here because DCR-G-Aff is significantly better than them according to Fig. \ref{fig:mse_vs_snr_n5}. For $k=1$, all of the methods have similar performance. For $k=6$ and $k=9$, the proposed method is consistently and significantly better than SPA, WDA, and DCR-G-Aff. More importantly, Fig. \ref{fig:mse_vs_snapshots_n5} also implies that a DNN model trained by the subspace representation learning approach on a specific number of snapshots can perform well across a wide range of unseen numbers of snapshots.

\subsubsection{On different SLAs} \label{sec:4_and_6_mra}
It is desirable to show that the main conclusions drawn from the $5$-element MRA experiments in Section \ref{sec:superior_performance_SNRs} in general hold true for an arbitrary $N$-element MRA. Here, we demonstrate that this is true for $4$-element and $6$-element MRAs.
Most of the hyperparameters stay the same as the setting in Section \ref{sec:numerical_results}. We find that $\delta=10^{-4}$ leads to unstable training in the DCR-G-Aff approach for the case of the $6$-element MRA so we increase $\delta$ to $10^{-3}$ in this particular case.
Results for the $4$-element MRA $\mathcal{S}=\{1,2,5,7\}$ are shown in Fig. \ref{fig:mse_vs_snr_n4}. Results for the $6$-element MRA $\mathcal{S}=\{1,2,5,6,12,14\}$ are shown in Fig. \ref{fig:mse_vs_snr_n6}. All of these results in Fig. \ref{fig:mse_vs_snr_n5}, \ref{fig:mse_vs_snr_n4}, and \ref{fig:mse_vs_snr_n6} demonstrate that the proposed method outperforms all of the baseline approaches. Furthermore, they seem to suggest our approach is consistently better than all baselines if $k\geq N$.
Although we do not include the results on the number of snapshots for the $4$-element and $6$-element MRAs in this paper (they can be found in \cite{chen2024deep}), our experiments show that they enjoy the same conclusion drawn from Fig. \ref{fig:mse_vs_snapshots_n5}.

\subsubsection{Other distances between subspaces} \label{sec:other_distances}
Although the geodesic distance, $d_k^{\text{Geo}}$, is the most natural choice for $d_k$ and is used to demonstrate the proposed methodology, other choices for $d_k$ are also possible. To study the effectiveness of subspace representation learning using different distances, we conduct an experiment under the setting of a $4$-element MRA, with the same setup as described in Section \ref{sec:4_and_6_mra}. Fig. \ref{fig:mse_vs_snr_n4_other_distances} shows that models trained using the distances in Table \ref{tab:subspace_distances} result in nearly identical performance for DoA estimation. This result aligns with theory, as $d_k^a \to 0$ implies $d_k^b \to 0$ when $d_k^a$ and $d_k^b$ are two different distances listed in Table \ref{tab:subspace_distances}. Therefore, we argue that using different distances between subspaces is likely to yield similar performance in our methodology.
\input{mse_snr_n4_other_distances}

\subsubsection{Random source powers} \label{sec:random_powers}
To relax the equal power assumption, new models are trained and evaluated with random source powers satisfying the condition $\frac{\max_i p_i}{\min_j p_j}\leq 10$. Except for the source power assumption, we follow the same setting as the 4-element MRA used in Section \ref{sec:4_and_6_mra}. Fig. \ref{fig:mse_vs_snr_n4_random_source_powers} shows that our approach significantly outperforms SPA, DCR-T, and DCR-G-Fro. Although the performance gap between DCR-G-Aff and our approach is greatly reduced compared to Fig. \ref{fig:mse_vs_snr_n4}, the relative ranking of these methods remains unchanged.
\input{mse_snr_n4_random_source_powers}
\subsection{Comparison to the proposed gridless end-to-end approach}  \label{sec:end2end}
    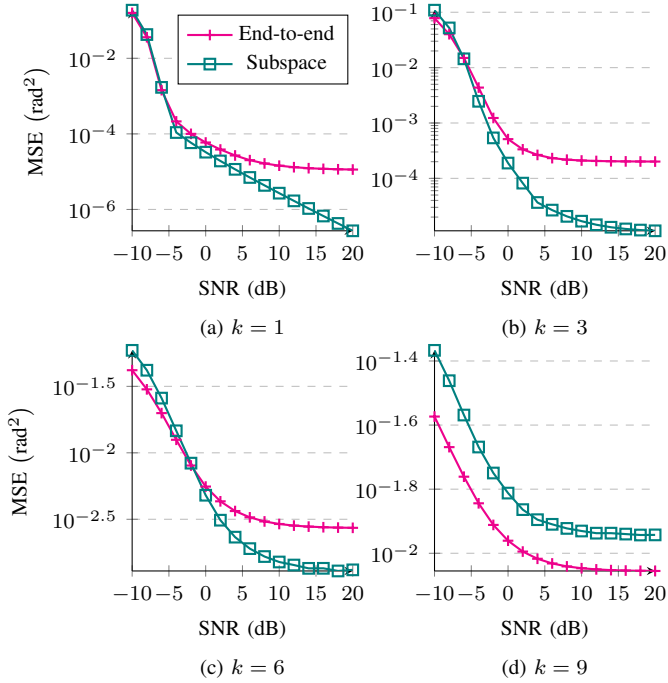
\begin{figure}[h]
        \centering
        \begin{tikzpicture}
        \begin{axis}[
            font=\footnotesize,
            width=4.5cm,
            height=4.5cm,
            axis lines = left,
            xlabel = SNR (dB),
            ylabel = MSE $\left(\text{rad}^2\right)$,
            legend style={anchor=north east},
            xtick={-10,-5,0,5,10,15,20},
            title={\footnotesize (a) $k=1$},
            title style={at={(0.5,-0.6)},anchor=south},
            xmax = 20,
            ymajorgrids=true,
            grid style=dashed,
            ymode=log
        ]

        \addplot [
            color=magenta,
            mark=+,
            thick
        ]
        coordinates {
        (-10,0.1591)
        (-8,0.036087)
        (-6,0.0014349)
        (-4,0.00021392)
        (-2,0.00010011)
        (0,5.9147e-05)
        (2,3.8589e-05)
        (4,2.6823e-05)
        (6,2.0162e-05)
        (8,1.6515e-05)
        (10,1.4402e-05)
        (12,1.3125e-05)
        (14,1.235e-05)
        (16,1.187e-05)
        (18,1.1567e-05)
        (20,1.1378e-05)
        };
    
        \addlegendentry{End-to-end}
            
        \addplot [
            color=teal,
            mark=square,
            thick
        ]
        coordinates {
        (-10,0.18772)
        (-8,0.042534)
        (-6,0.0017021)
        (-4,0.00010862)
        (-2,5.8248e-05)
        (0,3.2799e-05)
        (2,1.9206e-05)
        (4,1.1526e-05)
        (6,7.0214e-06)
        (8,4.333e-06)
        (10,2.6957e-06)
        (12,1.6873e-06)
        (14,1.0611e-06)
        (16,6.7072e-07)
        (18,4.2601e-07)
        (20,2.7219e-07)
        };   
        \addlegendentry{Subspace}
    
        \end{axis}

        \begin{axis}[
            font=\footnotesize,
            width=4.5cm,
            height=4.5cm,
            at={(4cm,0cm)},
            axis lines = left,
            xlabel = SNR (dB),
            xmax = 20,
            xtick={-10,-5,0,5,10,15,20},
            title={\footnotesize (b) $k=3$},
            title style={at={(0.5,-0.6)},anchor=south},
            ymajorgrids=true,
            grid style=dashed,
            ymode=log
        ]
    
        \addplot [
            color=magenta,
            mark=+,
            thick
        ]
        coordinates {
        (-10,0.078074)
        (-8,0.040158)
        (-6,0.014969)
        (-4,0.0043606)
        (-2,0.0012315)
        (0,0.0005111)
        (2,0.0003344)
        (4,0.00026513)
        (6,0.00023198)
        (8,0.00021765)
        (10,0.00021024)
        (12,0.00020616)
        (14,0.00020388)
        (16,0.00020253)
        (18,0.00020171)
        (20,0.00020121)
        };
            
        \addplot [
            color=teal,
            mark=square,
            thick
        ]
        coordinates {
        (-10,0.10898)
        (-8,0.052135)
        (-6,0.014429)
        (-4,0.0024658)
        (-2,0.00053826)
        (0,0.000189)
        (2,8.2042e-05)
        (4,3.6669e-05)
        (6,2.6712e-05)
        (8,2.0609e-05)
        (10,1.6855e-05)
        (12,1.454e-05)
        (14,1.3103e-05)
        (16,1.2209e-05)
        (18,1.165e-05)
        (20,1.1301e-05)
        };
        
        \end{axis}
    
        \begin{axis}[
            font=\footnotesize,
            width=4.5cm,
            height=4.5cm,
            at={(0cm,-4.5cm)},
            axis lines = left,
            xlabel = SNR (dB),
            ylabel = MSE $\left(\text{rad}^2\right)$,
            xmax = 20,
            xtick={-10,-5,0,5,10,15,20},
            title={\footnotesize (c) $k=6$},
            title style={at={(0.5,-0.6)},anchor=south},
            ymajorgrids=true,
            grid style=dashed,
            ytick={10^(-2.5),10^(-2),10^(-1.5),10^(-1)},
            ymode=log
        ]
        
        \addplot [
            color=magenta,
            mark=+,
            thick
        ]
        coordinates {
        (-10,0.04183)
        (-8,0.030029)
        (-6,0.019892)
        (-4,0.012533)
        (-2,0.0080243)
        (0,0.0055732)
        (2,0.0043165)
        (4,0.0036473)
        (6,0.00327)
        (8,0.0030466)
        (10,0.002914)
        (12,0.0028338)
        (14,0.0027851)
        (16,0.0027556)
        (18,0.0027376)
        (20,0.0027267)
        };
    
        \addplot [
            color=teal,
            mark=square,
            thick
        ]
        coordinates {
        (-10,0.059005)
        (-8,0.041805)
        (-6,0.025794)
        (-4,0.014643)
        (-2,0.0083602)
        (0,0.0047964)
        (2,0.0031122)
        (4,0.0023203)
        (6,0.0019022)
        (8,0.0016585)
        (10,0.0015088)
        (12,0.0014314)
        (14,0.0013516)
        (16,0.0013525)
        (18,0.0012918)
        (20,0.0013169)
        };
            
        \end{axis}
            
        \begin{axis}[
            font=\footnotesize,
            width=4.5cm,
            height=4.5cm,
            at={(4cm,-4.5cm)},
            axis lines = left,
            xlabel = SNR (dB),
            xmax = 20,
            xtick={-10,-5,0,5,10,15,20},
            title={\footnotesize (d) $k=9$},
            title style={at={(0.5,-0.6)},anchor=south},
            ymajorgrids=true,
            grid style=dashed,
            ytick={10^(-2),10^(-1.8),10^(-1.6),10^(-1.4)},
            ymode=log
        ]
        
        \addplot [
            color=magenta,
            mark=+,
            thick
        ]
        coordinates {
        (-10,0.026705)
        (-8,0.021439)
        (-6,0.017338)
        (-4,0.01432)
        (-2,0.01226)
        (0,0.010945)
        (2,0.010124)
        (4,0.0096127)
        (6,0.0092955)
        (8,0.0091005)
        (10,0.0089814)
        (12,0.0089074)
        (14,0.008861)
        (16,0.0088334)
        (18,0.0088169)
        (20,0.008807)
        };
            
        \addplot [
            color=teal,
            mark=square,
            thick
        ]
        coordinates {
        (-10,0.042967)
        (-8,0.034553)
        (-6,0.026997)
        (-4,0.021455)
        (-2,0.017802)
        (0,0.01541)
        (2,0.013684)
        (4,0.01273)
        (6,0.012312)
        (8,0.011956)
        (10,0.011735)
        (12,0.011554)
        (14,0.011555)
        (16,0.011479)
        (18,0.011403)
        (20,0.011412)
        };
        \end{axis}
    \end{tikzpicture}
    \caption{MSE vs. SNR. $N=5$. $M=10$. For $k\in\{1,3,6\}$, the performance of the gridless end-to-end approach saturates at a higher MSE than the subspace representation learning method as the SNR increases. For $k=9$, the gridless end-to-end approach shows consistently better performance.}
    \label{fig:end2end_mse_vs_snr_n5}
\end{figure}
To answer the question posed in Section \ref{sec:gridless_end2end_training}, we use the same WRN-16-8 but replace the final affine layer by $M-1$ affine heads whose number of output neurons are $1,2,3,\cdots,M-1$, as illustrated in Fig. \ref{fig:gridless_end2end_architecture}. The squared loss functions of different dimensions are adopted as shown in (\ref{eq:squared_loss_optimal_permutation}). All of the settings here are the same as the ones described in Section \ref{sec:settings} and \ref{sec:training}. The best learning rate is $0.2$ according to a simple grid search. Fig. \ref{fig:end2end_mse_vs_snr_n5} shows that the gridless end-to-end approach tends to saturate its performance earlier than the subspace representation learning approach for $k\in\{1,3,6\}$ as the SNR increases. As a result, the subspace representation learning approach shows significantly better performance at high SNRs. However, for $k=9$, it is consistently worse than the gridless end-to-end approach. Although the gridless end-to-end approach does not have a grid at the output layer, its behavior of hitting an early plateau seems to be similar to grid-based methods that are limited by their grid resolution. Overall, subspace representation learning gives better performance than the gridless end-to-end approach and we can deduce that learning subspace representations is more beneficial than learning angles directly.

\subsection{Robustness to array imperfections} \label{sec:robustness2imperfections}
\noindent With regard to array imperfections, we use the imperfect array manifold introduced by Liu \textit{et al.} \cite{liu2018direction}. The exact formulation is given below. Let the degree of imperfections be controlled by a scalar $\rho\in[0,1]$. A larger $\rho$ makes the imperfections more severe and $\rho=0$ means the array is perfect. Define the following real hyperparameters
\begin{equation}
    e_1,\cdots,e_M,g_1,\cdots,g_M,h_1,\cdots,h_M
\end{equation}
and a complex hyperparameter $\gamma$. The array manifold with sensor position errors is given by $\mathbf{a}_{\rho}(\theta):[0,\pi]\to\mathbb{C}^M$ such that
\begin{equation}
    [\mathbf{a}_{\rho}(\theta)]_i=e^{j2\pi \left(i-1-\frac{(M-1)}{2}+\rho e_i\right)\frac{d}{\lambda}\cos\theta}
\end{equation}
for $i\in[M]$. Then, an imperfect array manifold $\tilde{\mathbf{a}}_{\rho}(\theta)$ of an $M$-element ULA can be defined by
\begin{equation}
    \tilde{\mathbf{a}}_{\rho}(\theta)=\mathbf{C}_{\rho}\mathbf{G}_{\rho}\mathbf{H}_{\rho}\mathbf{a}_{\rho}(\theta)
\end{equation}
where the gain bias is modeled by
\begin{equation}
    \mathbf{G_{\rho}}=\mathbf{I}+\rho\text{diag}\left(g_1,g_2,\cdots,g_M\right),
\end{equation}
the phase bias is modeled by
\begin{equation}
    \mathbf{H}_{\rho}=\text{diag}\left(e^{j\rho h_1},e^{j\rho h_2},\cdots,e^{j\rho h_M}\right),
\end{equation}
and the intersensor mutual coupling is modeled by
\begin{equation}
    \mathbf{C}_{\rho}=\mathbf{I}+\rho\text{Toep}\left(\begin{bmatrix}
    0&\gamma&\gamma^2&\cdots&\gamma^{M-1}
    \end{bmatrix}^{\mathsf{T}}\right).
\end{equation}
For the hyperparameters, we use $e_1=0,e_2=\cdots=e_6=-0.2,e_7=\cdots=e_{10}=0.2$, $g_1=0,g_2=\cdots=g_6=0.2,g_7=\cdots=g_{10}=-0.2$, and $h_1=0,h_2=\cdots=h_6=-\frac{1}{6}\pi,h_7=\cdots=h_{10}=\frac{1}{6}\pi$. To train a model for imperfect arrays, we uniformly select $\rho$ at random on the unit interval $[0,1]$. To train a model for a perfect array, we use $\rho=0$.

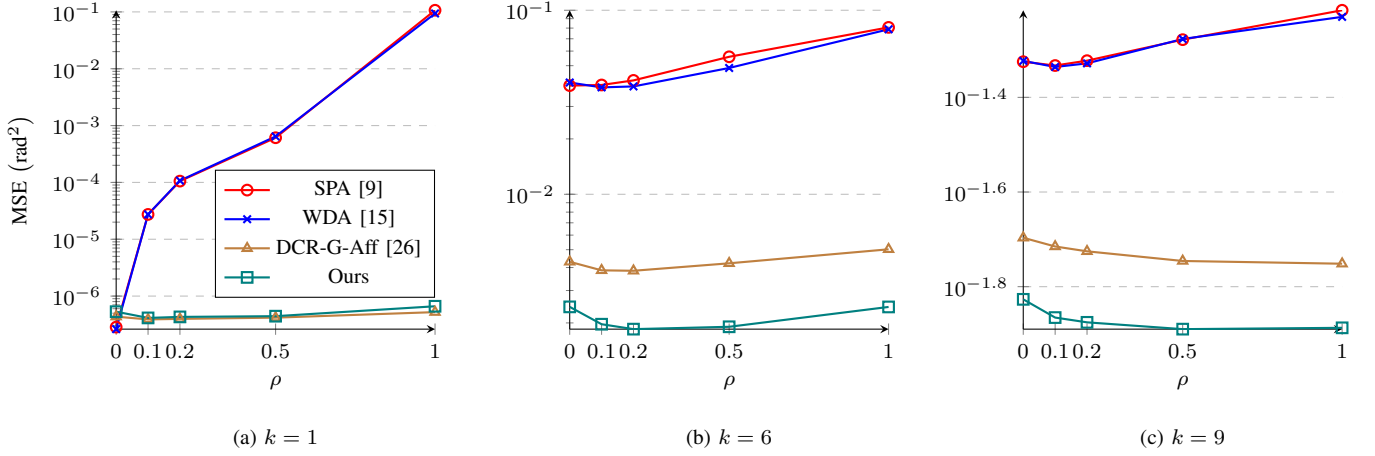
\begin{figure*}[ht]
\centering
\begin{tikzpicture}
\begin{axis}[
    font=\footnotesize,
    width=5.8cm,
    height=5.8cm,
    axis lines = left,
    xlabel = $\rho$,
    ylabel = MSE $\left(\text{rad}^2\right)$,
    legend style={at={(1.0,0.5)},anchor=north east},
    xtick={0,0.1,0.2,0.5,1.0},
    title={\footnotesize (a) $k=1$},
    title style={at={(0.5,-0.45)},anchor=south},
    xmax = 1.0,
    ymajorgrids=true,
    grid style=dashed,
    ymode=log
]
\addplot [
    color=red,
    mark=o,
    thick,
]
coordinates {(0.0,2.8455e-07)
(0.1,2.7333e-05)
(0.2,0.00010538)
(0.5,0.00061089)
(1.0,0.10687)
};

\addlegendentry{SPA \cite{yang2014discretization}}

\addplot [
    color=blue,
    mark=x,
    thick
    ]
    coordinates {(0.0,2.6211e-07)
    (0.1,2.7378e-05)
    (0.2,0.00010766)
    (0.5,0.00064433)
    (1.0,0.093829)
    };

\addlegendentry{WDA \cite{wang2019grid}}

\addplot [
    color=brown,
    mark=triangle,
    thick
    ]
    coordinates {(0.0,4.3474e-07)
    (0.1,3.8859e-07)
    (0.2,3.974e-07)
    (0.5,4.1758e-07)
    (1.0,5.2226e-07)
    };

\addlegendentry{DCR-G-Aff \cite{barthelme2021doa}}

\addplot [
    color=teal,
    mark=square,
    thick
    ]
    coordinates
    {(0.0,5.3135e-07)
    (0.1,4.1297e-07)
    (0.2,4.3023e-07)
    (0.5,4.4297e-07)
    (1.0,6.6199e-07)
    };
\addlegendentry{Ours}
\end{axis}

\begin{axis}[
    font=\footnotesize,
    width=5.8cm,
    height=5.8cm,
    at={(6cm,0cm)},
    axis lines = left,
    xlabel = $\rho$,
    xmax = 1.0,
    xtick={0,0.1,0.2,0.5,1.0},
    title={\footnotesize (b) $k=6$},
    title style={at={(0.5,-0.45)},anchor=south},
    ymajorgrids=true,
    grid style=dashed,
    ymax= 1e-1,
    ymode=log
]

\addplot [
    color=red,
    mark=o,
    thick
]
coordinates {(0.0,0.038966)
(0.1,0.039269)
(0.2,0.041583)
(0.5,0.055866)
(1.0,0.080715)
};

\addplot [
    color=blue,
    mark=x,
    thick
    ]
    coordinates {(0.0,0.040492)
    (0.1,0.0381)
    (0.2,0.038562)
    (0.5,0.048571)
    (1.0,0.078805)
    };

\addplot [
    color=brown,
    mark=triangle,
    thick
    ]
    coordinates {(0.0,0.0042973)
    (0.1,0.0038622)
    (0.2,0.0038408)
    (0.5,0.0042169)
    (1.0,0.0050243)
    };

\addplot [
    color=teal,
    mark=square,
    thick
    ]
    coordinates {(0.0,0.002445)
    (0.1,0.001968)
    (0.2,0.0018504)
    (0.5,0.0019057)
    (1.0,0.0024426)
    };
\end{axis}

\begin{axis}[
    font=\footnotesize,
    width=5.8cm,
    height=5.8cm,
    at={(12cm,0cm)},
    axis lines = left,
    xlabel = $\rho$,
    xmax = 1.0,
    xtick={0,0.1,0.2,0.5,1.0},
    title={\footnotesize (c) $k=9$},
    title style={at={(0.5,-0.45)},anchor=south},
    ymajorgrids=true,
    grid style=dashed,
    ymode=log
]

\addplot [
    color=red,
    mark=o,
    thick
]
coordinates {(0.0,0.047352)
(0.1,0.046479)
(0.2,0.047594)
(0.5,0.052714)
(1.0,0.060822)
};

\addplot [
    color=blue,
    mark=x,
    thick
    ]
    coordinates {(0.0,0.047545)
    (0.1,0.046156)
    (0.2,0.046971)
    (0.5,0.052887)
    (1.0,0.058894)
    };

\addplot [
    color=brown,
    mark=triangle,
    thick
    ]
    coordinates {(0.0,0.020108)
    (0.1,0.019275)
    (0.2,0.018827)
    (0.5,0.017962)
    (1.0,0.017722)
    };

\addplot [
    color=teal,
    mark=square,
    thick
    ]
    coordinates {(0.0,0.014904)
    (0.1,0.013638)
    (0.2,0.013325)
    (0.5,0.012891)
    (1.0,0.012974)
    };
\end{axis}

\end{tikzpicture}
\caption{MSE vs. the array imperfection parameter $\rho$. Note that only one DNN model is trained for our approach. Unlike model-based methods that give significantly worse MSE as $\rho$ increases, our approach is robust to array imperfections without being given $\rho$ or the knowledge about the degree of imperfections.}
\label{fig:mse_vs_rho_n5}
\end{figure*}
Fig. \ref{fig:mse_vs_rho_n5} shows the MSE in terms of the array imperfection parameter $\rho$ for different numbers of sources. Indeed, a different $\rho$ represents a different array; thus, one can collect a new dataset and then specifically train a new model. However, here, we train a single model on a joint dataset collected from different arrays. The MSE of SPA and WDA both get worse as $\rho$ increases, verifying that both methods suffer from model mismatch and are not robust to array imperfections. Even though no attempts have been made to SPA and WDA to contend with array imperfections, such corrections are nontrivial and require the knowledge of imperfections. On the other hand, the MSE of the proposed method stays at the same level despite the increasing degree of imperfections, implying that the subspace representation learning approach is robust to array imperfections.
Although we do not include the results for the $4$-element and $6$-element MRAs here (they can be found in \cite{chen2024deep}), our experiments show that they enjoy the same conclusion drawn from Fig. \ref{fig:mse_vs_rho_n5}.

\subsection{Consistent rank sampling} \label{sec:consistent_rank_sampling_experiment}
\noindent To study the speedup and potential performance regression of consistent rank sampling, two subspace representation learning models are trained with and without consistent rank sampling. This study is conducted on a 4-element MRA with the same setup used in Section \ref{sec:4_and_6_mra}. The model trained without consistent rank sampling achieves an empirical risk of $0.21312$ on the validation set and a training speed of $356.67$ seconds per epoch. On the other hand, the model trained with consistent rank sampling achieves an empirical risk of $0.21322$ and a training speed of $172.30$ seconds per epoch. As a result, consistent rank sampling provides about a $2\times$ training speedup with negligible performance regression. The study is run on an NVIDIA RTX 4090 GPU.
\section{Conclusion}
\noindent A new methodology learning subspace representations is proposed for robust estimation of more sources than sensors. To learn subspace representations, the codomain of a DNN model, is defined as a union of Grassmannians reflecting signal subspaces of different dimensions. Then, a family of loss functions is proposed as functions of the principal angles between subspaces to ensure rational invariance. In particular, we use geodesic distances on Grassmannians to train a DNN model and prove that it is possible for a ReLU network to approximate signal subspaces. Because a subspace is invariant to the selection of the basis, our methodology expands the solution space of a DNN model compared to existing approaches that learn covariance matrices. In addition, due to its geometry-agnostic nature, our methodology is robust to array imperfections. To study the possibility of bypassing the root-MUSIC algorithm, we propose a gridless end-to-end approach that directly learns a mapping from sample SCMs to DoAs. Numerical results show that subspace representation learning outperforms existing SDP-based approaches including the SPA and WDA, DNN-based covariance matrix reconstruction methods, and the gridless end-to-end approach under the standard assumptions. These results imply that learning subspace representations is more beneficial than learning covariance matrices or angles directly.
\section*{Acknowledgments}
\noindent This work was supported in part by IEEE Signal Processing Society Scholarship, and in part by NSF under Grant CCF-2225617 and Grant CCF-2124929. The authors would like to thank Dr. Ching-Hua Lee for his discussions on the gridless end-to-end training and Mr. Rushabha Balaji for his suggestions and comments regarding consistent rank sampling.

{
\appendices
\section{Proof of Theorem \ref{thm:subspace_approximation}} \label{proof:subspace_approximation}
\begin{proof}
Let $\mathcal{I}=\left\{(i,j)\in[k]\times[k] \mid i\neq j\right\}$.
For every $(i,j)\in\mathcal{I}$, define
\begin{equation}
\mathcal{F}_{i,j}=\left\{\boldsymbol{\theta}\in[0,\pi]^k \mid \theta_i=\theta_j\right\}.
\end{equation}
Pick $\delta>0$ and let $\mu$ denote the Lebesgue measure. Because $\mathcal{F}_{i,j}$ is closed and $\mu(\mathcal{F}_{i,j})=0$ for every $(i,j)\in\mathcal{I}$, there exists an open set
$
    \mathcal{F}_{\delta}\supset\bigcup_{(i,j)\in\mathcal{I}}\mathcal{F}_{i,j}
$
such that
$
    \mu(\mathcal{F}_{\delta})<\delta.
$
Therefore, $\mathcal{E}_{\delta}=[0,\pi]^k\setminus\mathcal{F}_{\delta}$ is compact.
Now, note that $\mathbf{A}(\boldsymbol{\theta})$ is a rank-$k$ matrix for every $\boldsymbol{\theta}\in\mathcal{E}_{\delta}$ due to the Vandermonde structure. It follows that the function $\mathbf{X}\mapsto P_{\mathbf{X}}$ is continuous on $\mathbf{A}(\mathcal{E}_{\delta})$. On the other hand, the mapping $\mathbf{R}_{\mathcal{S}}\mapsto\mathbf{R}_0$ is affine on $\mathbf{A}(\mathcal{E}_{\delta})$ since the SLA has no holes in its co-array. As $\mathbf{R}_0=\mathbf{A}(\boldsymbol{\theta})\mathbf{P}\mathbf{A}^{\mathsf{H}}(\boldsymbol{\theta})$, we have $P_{\mathbf{R}_0}=P_{\mathbf{A}(\boldsymbol{\theta})}$, implying that $\mathbf{R}_{\mathcal{S}}\mapsto P_{\mathbf{A}(\boldsymbol{\theta})}$ is continuous on $\mathbf{A}(\mathcal{E}_{\delta})$. By Theorem 1 of \cite{chen2022improved}, any continuous piecewise linear function can be represented by a ReLU network. Because the set of continuous piecewise linear functions is dense in the set of continuous functions on any compact subset of $\mathbb{C}^{N\times N}$, it follows that, for every $\epsilon$, there is a ReLU network $f$ such that
\begin{equation}
\sup_{\boldsymbol{\theta}\in\mathcal{E}_{\delta}} \left\lVert f\left(\mathbf{R}_{\mathcal{S}}\right) - P_{\mathbf{A}(\boldsymbol{\theta})} \right\rVert_F < \epsilon.
\end{equation}
Note that $\mathbf{R}_{\mathcal{S}}(\mathcal{E}_{\delta})$ is still compact since $\boldsymbol{\theta}\mapsto\mathbf{R}_{\mathcal{S}}$ is continuous. By Lemma \ref{lemma:geodesic_upper_bound},
\begin{equation}
    \int_{\mathcal{E}_{\delta}}d_k^{\text{Geo}}\left( f\left(\mathbf{R}_{\mathcal{S}}\right),P_{\mathbf{A}(\boldsymbol{\theta})}\right) d\boldsymbol{\theta}< \pi^k\sqrt{k}\sin^{-1}\left(\frac{\epsilon}{\sqrt{2}}\right).
\end{equation}
As $f$ is continuous and every nonzero orthogonal projection is bounded, there exists $L>0$ such that $\left\lVert f\left(\mathbf{R}_{\mathcal{S}}\right) - P_{\mathbf{A}(\boldsymbol{\theta})} \right\rVert_F<L$ for every $\boldsymbol{\theta}\in\mathcal{F}_{\delta}$, which implies
\begin{equation}
    \int_{\mathcal{F}_{\delta}}d_k^{\text{Geo}}\left( f\left(\mathbf{R}_{\mathcal{S}}\right),P_{\mathbf{A}(\boldsymbol{\theta})}\right) d\boldsymbol{\theta}<\delta \sqrt{k}\sin^{-1}\left(\frac{L}{\sqrt{2}}\right).
\end{equation}
The claim is proved because both $\delta>0$ and $\epsilon>0$ can be arbitrarily small, and $\sin^{-1}(x)\to 0$ as $x\to 0^+$.
\end{proof}

\section{Proof of Lemma \ref{lemma:geodesic_upper_bound}} \label{proof:geodesic_upper_bound}
\begin{proof}
Because there is a one-to-one correspondance between the set of linear subspaces and the set of orthogonal projectors, a distance $d:\text{Gr}(k,M)\times \text{Gr}(k,M)\to[0,\infty)$ between $\mathcal{U}_1\in\text{Gr}(k,M)$ and $\mathcal{U}_2\in\text{Gr}(k,M)$ can be defined as
\begin{equation}
    d(\mathcal{U}_1,\mathcal{U}_2)=\left\lVert P_{\mathcal{U}_1}-P_{\mathcal{U}_2} \right\rVert_F
\end{equation}
where $P_{\mathcal{U}_1}$ and $P_{\mathcal{U}_2}$ are the orthogonal projectors onto $\mathcal{U}_1$ and $\mathcal{U}_2$, respectively. Then, it follows that
$
        d^2(\mathcal{U}_1,\mathcal{U}_2)=2k-2\text{tr}\left(P_{\mathcal{U}_1}P_{\mathcal{U}_2}\right)
$ which is equivalent to
\begin{equation}
    \begin{split}
    2k-2\sum_{i=1}^k\sigma_i^2\left(P_{\mathcal{U}_1}P_{\mathcal{U}_2}\right)&=2k-2\sum_{i=1}^k\sigma_i^2\left(\mathbf{U}_1^{\mathsf{H}}\mathbf{U}_2\right)\\
    &=2k-2\sum_{i=1}^k\cos^2\phi_i
    \end{split}
\end{equation}
where $\mathbf{U}_1$ and $\mathbf{U}_2$ are matrices whose columns form unitary bases of $\mathcal{U}_1$ and $\mathcal{U}_2$. Therefore, we have
\begin{equation} \label{eq:principal_angle_projector_relation}
    \left\lVert P_{\mathcal{U}_1}-P_{\mathcal{U}_2} \right\rVert_F=\sqrt{2}\left(\sum_{i=1}^k\sin^2\phi_i\right)^{\frac{1}{2}}
\end{equation}
which was shown in \cite{stewart1991perturbation}. Finally, (\ref{eq:principal_angle_projector_relation}) implies that
\begin{equation}
    \phi_i\leq\sin^{-1}\left(\frac{\left\lVert P_{\mathcal{U}_1}-P_{\mathcal{U}_2} \right\rVert_F}{\sqrt{2}}\right)
\end{equation}
for every $i\in[k]$.
\end{proof}
\section{Learning rates} \label{sec:learning_rates_grid_search}
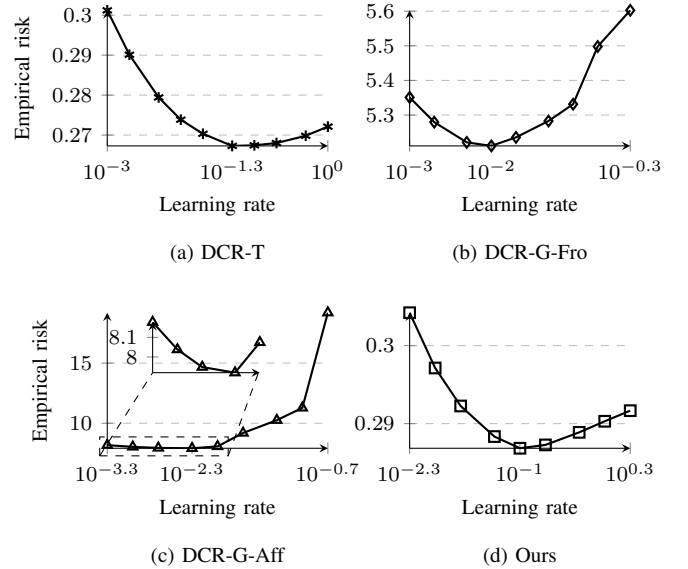
\begin{figure}[h]
\centering
\begin{tikzpicture}

\begin{axis}[
    at={(0cm,0cm)},
    font=\footnotesize,
    width=4.5cm,
    height=3.375cm,
    axis lines = left,
    xlabel = Learning rate,
    ylabel = Empirical risk,
    legend style={anchor=north east},
    xtick={0.001,0.05,1.0},
    title style={at={(0.5,-0.45)},anchor=south},
    ymajorgrids=true,
    grid style=dashed,
    xmode=log,
    title={\footnotesize (a) DCR-T},
    title style={at={(0.5,-1.05)},anchor=south}
]
\addplot [
    color=black,
    mark=asterisk,
    thick
]
coordinates {
    (0.001,0.30122)
    (0.002,0.29011)
    (0.005,0.27942)
    (0.01,0.27385)
    (0.02,0.27032)
    (0.05,0.26729)
    (0.1,0.26743)
    (0.2,0.26801)
    (0.5,0.26982)
    (1.0,0.27212)};
\end{axis}

\begin{axis}[
    at={(4cm,0cm)},
    font=\footnotesize,
    width=4.5cm,
    height=3.375cm,
    axis lines = left,
    xlabel = Learning rate,
    legend style={anchor=north east},
    xtick={0.001,0.01,0.5},
    title style={at={(0.5,-0.45)},anchor=south},
    ymajorgrids=true,
    grid style=dashed,
    xmode=log,
    title={\footnotesize (b) DCR-G-Fro},
    title style={at={(0.5,-1.05)},anchor=south}
]
\addplot [
    color=black,
    mark=diamond,
    thick
]
coordinates {
    (0.001,5.351)
    (0.002,5.2796)
    (0.005,5.2212)
    (0.01,5.211)
    (0.02,5.2356)
    (0.05,5.2829)
    (0.1,5.3314)
    (0.2,5.4979)
    (0.5,5.6024)};
\end{axis}

\begin{axis}[
    at={(0cm,-4cm)},
    font=\footnotesize,
    width=4.5cm,
    height=3.375cm,
    axis lines = left,
    xlabel = Learning rate,
    ylabel = Empirical risk,
    legend style={anchor=north east},
    xmin=0.0005,
    xmax=0.2,
    xtick={0.0005,0.005,0.2},
    title style={at={(0.5,-0.45)},anchor=south},
    ymajorgrids=true,
    grid style=dashed,
    xmode=log,
    title={\footnotesize (c) DCR-G-Aff},
    title style={at={(0.5,-1.05)},anchor=south}
]
\addplot [
    color=black,
    mark=triangle,
    thick
]
coordinates {
    (0.0005,8.1778)
    (0.001,8.0379)
    (0.002,7.9482)
    (0.005,7.9196)
    (0.01,8.0743)
    (0.02,9.1904)
    (0.05,10.2533)
    (0.1,11.2798)
    (0.2,19.1912)};

\end{axis}

\begin{axis}[
    at={(0.6cm,-3cm)},
    font=\footnotesize,
    width=3cm,
    height=2.25cm,
    axis lines = left,
    xmin=0.0005,
    xmax=0.01,
    xtick={0.0001},
    title style={at={(0.5,-0.45)},anchor=south},
    xmode=log
]
\addplot [
    color=black,
    mark=triangle,
    thick
]
coordinates {
    (0.0005,8.1778)
    (0.001,8.0379)
    (0.002,7.9482)
    (0.005,7.9196)
    (0.01,8.0743)};

\end{axis}

\draw[draw=black,dashed] (-0.1,-4.1) rectangle ++(1.7,0.25);
\draw [dashed] (-0.1,-4.1) -- (0.6,-3);
\draw [dashed] (1.6,-4.1) -- (2.0,-3);

\begin{axis}[
    at={(4cm,-4cm)},
    scaled y ticks=false,
    font=\footnotesize,
    width=4.5cm,
    height=3.375cm,
    axis lines = left,
    xlabel = Learning rate,
    legend style={anchor=north east},
    scaled ticks=false,
    xmin=0.005,
    xmax=2.0,
    xtick={0.005,0.1,2.0},
    ytick={0.29,0.30},
    ymajorgrids=true,
    grid style=dashed,
    xmode=log,
    title={\footnotesize (d) Ours},
    title style={at={(0.5,-1.05)},anchor=south}
]
\addplot [
    color=black,
    mark=square,
    thick
]
coordinates {
    (0.005,0.30422)
    (0.01,0.29713)
    (0.02,0.29226)
    (0.05,0.28834)
    (0.1,0.28684)
    (0.2,0.28727)
    (0.5,0.28888)
    (1.0,0.2903)
    (2.0,0.29166)};
\end{axis}

\end{tikzpicture}
\caption{Search of the best learning rates. Empirical risk on the validation set vs. the maximum learning rate in the one-cycle learning rate scheduler.}
\label{fig:lr_search_for_each_method}
\end{figure}
\noindent To determine the best learning rates to use in Section \ref{sec:numerical_results}, a grid search of the best maximum learning rate in the one-cycle learning rate scheduler \cite{smith2019super} is performed for each approach. For each $k\in[M-1]$, there are $2\times 10^6$ and $6\times 10^5$ random data points for training and validation, respectively, leading to a training dataset of size $L_{\text{train}}=9\times 2\times 10^6$ and a validation dataset of size $L_{\text{val}}=9\times 6\times 10^5$. Figure \ref{fig:lr_search_for_each_method} shows that the best learning rates for DCR-T, DCR-G-Fro, DCR-G-Aff, and the proposed approach are $0.05$, $0.01$, $0.005$, and $0.1$, respectively. These learning rates are also used to train all the corresponding models in Section \ref{sec:4_and_6_mra}, \ref{sec:other_distances}, \ref{sec:random_powers}, \ref{sec:end2end}, and \ref{sec:consistent_rank_sampling_experiment}.
\section{Invariance-aware loss functions} \label{sec:invariance_aware_losses}
\noindent There are other loss functions that deviate from covariance reconstruction for gridless DoA estimation \cite{chen2025comparative}. These loss functions are developed based on the scale-invariant signal-to-distortion ratio (SI-SDR) to partially address the primary issue identified in Section \ref{sec:subspace_representation_learning}. Although the new loss functions proposed in \cite{chen2025comparative} underperform subspace representation learning, they provide evidence that loss functions with greater degrees of invariance can achieve better DoA estimation performance.
}

\bibliographystyle{IEEEtran}
\bibliography{ref}

\vfill

\end{document}